\documentclass[pra,aps,twocolumn,floatfix,showpacs]{revtex4}
\usepackage{graphicx,psfrag,amsmath,calc,amssymb}
\begin{document}
\title{Fermi-Fermi Mixtures in the Strong Attraction Limit}
\author{M. Iskin and C. A. R. S{\'a} de Melo}
\affiliation{School of Physics, Georgia Institute of Technology, Atlanta, Georgia 30332-0432, USA. \\
Joint Quantum Institute, NIST and University of Maryland, College Park, Maryland 20742-4111, USA.}

\date{\today}

\begin{abstract}

The phase diagrams of low density Fermi-Fermi mixtures with equal or unequal masses and equal
or unequal populations are described at zero and finite temperatures in the strong
attraction limit. In this limit, the Fermi-Fermi mixture can be described by 
a weakly interacting Bose-Fermi mixture, where the bosons correspond to Feshbach 
molecules and the fermions correspond to excess atoms. 
First, we discuss the three and four fermion scattering processes, and use the
exact boson-fermion and boson-boson scattering lengths to generate the phase diagrams 
in terms of the underlying fermion-fermion scattering length.
In three dimensions, in addition to the normal and uniform superfluid phases,
we find two stable non-uniform states corresponding to
(1) phase separation between pure unpaired (excess)
and pure paired fermions (molecular bosons); and 
(2) phase separation between pure excess fermions 
and a mixture of excess fermions and molecular bosons. 
Lastly, we also discuss the effects of the trapping potential in the density profiles of 
condensed and non-condensed molecular bosons, and excess fermions at zero and finite 
temperatures, and discuss possible implications of our findings to experiments
involving mixtures of ultracold fermions.

\end{abstract}
\maketitle

\section{Introduction}
\label{introduction}

The lack of precise control over standard condensed matter systems has hindered
the development of experiments that can probe systematically the effects
of strong correlations. However, the large degree of control in atomic systems, has
made them powerful tools for studying many condensed matter phenomena, 
and in particular novel superfluid phases~\cite{bourdel1,regal2,chin,partridge1,kinast2,zwierlein3}.
For instance, a current research frontier is the study of fermion mixtures with population 
imbalance~\cite{mit,rice,mit-2,rice-2}. 
Since the population of each component as well as their interaction strength
are experimentally tunable, these knobs enabled the
study of the BCS to BEC evolution in population imbalanced two-component fermion
superfluids~\cite{mit,rice,mit-2,rice-2}.
In contrast with the crossover physics found in the population balanced 
case~\cite{leggett,nsr,carlos}, these experiments have demonstrated the 
existence of phase transitions between normal and superfluid phases, 
as well as phase separation between superfluid (paired) and normal (excess) 
fermions as a function of 
population imbalance~\cite{liu,bedaque,sedrakian,castorina,carlson,pao,sheehy}.

Motivated by these recent experiments, there has been intense theoretical
interest in understanding the phase diagram of population
imbalanced mixtures~\cite{torma,pieri,yi,silva,haque,iskin-mixture,lobo,liu-mixture,mizushima}.
So far, an accurate description of such mixtures is only available 
in the weak fermion attraction limit, and it is yet to be 
developed for intermediate fermion attraction around unitarity, or
for the strong attraction limit. Some progress has been made 
in the strong attraction limit, where Fermi-Fermi mixtures were described
as a weakly interacting Bose-Fermi mixture~\cite{pieri, iskin-mixture, iskin-mixture2, taylor},
however, the effective boson-fermion and boson-boson scattering parameters were
obtained only in the Born approximation.
Strictly speaking, the Bose-Fermi description is valid only in the strong attraction limit, 
but may also provide semi-quantitative understanding of the phase diagram close to unitarity. 
Thus, the main goal of this manuscript is to analyze the boson-fermion
and boson-boson scattering parameters beyond the Born approximation
for arbitrary mass ratio of fermions, and use the effective Bose-Fermi mixture description
to generate improved phase diagrams and density profiles of Fermi-Fermi 
mixtures with equal or unequal masses in the strong 
attraction limit beyond the Born and mean-field 
approximations~\cite{iskin-mixture,pao-mixture,iskin-mixture2,duan-mixture,parish}. 

The main results of this manuscript are as follows.
First, we analyze three- and four-fermion scattering 
processes and obtain the exact boson-fermion and boson-boson scattering lengths 
as a function of mass anisotropy. Second, we use the exact boson-fermion
and boson-boson scattering parameters to construct the phase
diagram for Fermi-Fermi mixtures in the strong attraction limit.
In addition to the normal (N) and uniform superfluid (U) phases,
we find two different non-uniform phase separated (PS) states: 
(1) phase separation between pure unpaired (excess) 
and pure paired fermions (molecular bosons), and
(2) phase separation between pure excess fermions 
and a mixture of excess fermions and molecular bosons, 
depending on the fermion-fermion scattering parameter.
The phase boundaries are very sensitive to the masses of the fermions, and also to
the boson-fermion and boson-boson interactions. 
For equal mass mixtures, our results for the phase boundary between the PS(2) and 
the uniform (U) phase improves on previous saddle-point 
(mean-field) results, and the quantitative changes are substantial, but not
dramatic.
However, there is a dramatic increase in quantitative differences between 
mean-field and the present results for unequal mass mixtures as the mass ratio
deviates from one. In particular, these differences are more pronounced when 
heavier fermions are in excess indicating the importance of taking into account
scattering processes beyond the Born approximation.
Furthermore, we discuss the effects of the trapping potential on the density 
profiles of condensed and non-condensed molecular bosons, as well as excess 
fermions at zero and finite temperatures. Lastly, we discuss the implications 
of our findings to possible experiments involving Fermi-Fermi mixtures 
with equal or unequal masses and equal or unequal populations.

The remainder of this manuscript is organized as follows.
In Sec.~\ref{sec:hamiltonian}, we discuss briefly the Hamiltonian for Fermi-Fermi 
mixtures with equal or unequal masses and emphasize 
that the system reduces effectively to a Bose-Fermi mixture 
of molecular bosons and excess fermions in the strong 
attraction limit.
In Sec.~\ref{sec:a_BF}, we analyze the exact boson-fermion and boson-boson 
scattering lengths as a function of mass anisotropy, which are used to 
calculate the resulting phase diagram of Fermi-Fermi mixtures in 
the strong attraction limit. 
In Sec.~\ref{sec:BFmixture}, we discuss the stability of the effective
Bose-Fermi mixture of molecular bosons and excess fermions in three, two and one dimensions.
We analyze the stability of population imbalanced Fermi-Fermi mixtures 
in the strong attraction limit in Sec.~\ref{sec:FFmixture}, and 
we construct the phase diagrams of these systems in Sec.~\ref{sec:phase.diagram}.
In Sec.~\ref{sec:trap}, we discuss the effects of harmonic traps on the density 
profiles of condensed and non-condensed molecular bosons, and excess fermions 
at zero and finite temperatures, and show their experimental signatures.
Lastly, we present a summary of our conclusions in Sec.~\ref{sec:conclusions}.

\section{Hamiltonian}
\label{sec:hamiltonian}

In order to calculate the correct phase diagrams of Fermi-Fermi mixtures
in the strong attraction limit, it is necessary to obtain first the correct
scattering parameters between two Bose molecules (paired fermions), 
and also between a Bose molecule and an unpaired fermion. To achieve this task,
we begin by describing the Hamiltonian density for a mixture of fermions 
(in units of $\hbar = k_B = 1$) as 
\begin{eqnarray}
\label{eqn:hamiltonian}
H (x) &=& \sum_\sigma \bar \psi_{\sigma} (x) \left[ - \frac {\nabla^2}{ 2m_{\sigma} }  - \mu_{\sigma} \right] 
\psi_{\sigma} \nonumber \\
&-& g \bar \psi_{\uparrow} (x) \bar \psi_{\downarrow} (x) \psi_{\downarrow} (x) \psi_{\uparrow} (x),
\end{eqnarray}
where $\bar \psi_{\sigma} (x)$ is the field corresponding to the creation
of a fermion with pseudospin index $\sigma$, at position and time $x \equiv ({\bf r}, \tau)$.
Here, $g > 0$ is the strength of the attractive fermion-fermion interaction 
and $\sigma$ identifies two types of ($\uparrow$ and $\downarrow$) 
fermions. This notation allows the analysis of a mixture of fermions with 
equal or unequal masses, as well as equal or unequal chemical potentials.
Throughout the manuscript, we assume that the lighter fermions are always $\uparrow$ and
that the heavier fermions are always $\downarrow$, as intuitively suggested by the direction of the arrows,
and that the chemical potentials $\mu_{\sigma}$ fix the population (density) $n_{\sigma}$ 
of each type of fermion independently.

The contact interaction Hamiltonian given in Eq.~(\ref{eqn:hamiltonian})
generalizes the equal mass and equal chemical potential Hamiltonian that is used to study 
the BCS to BEC evolution within the functional integral formalism~\cite{carlos}.
The functional integral formulation~\cite{carlos,iskin-mixture,iskin-mixture2,taylor} captures 
some essential features of the evolution from BCS to BEC superfluidity for Fermi mixtures 
with equal or unequal masses as well as with equal or unequal populations. 
However, truly quantitative results are currently possible only in the BCS limit, 
where the theory is simple, but the temperatures required to reach the BCS regime 
are very low and hard to be achieved experimentally.
In the unitarity regime, experiments can be performed and the phase diagram 
can be explored since the critical temperature for superfluidity is attainable,
but an accurate theoretical description of this regime is still lacking.
While in the BEC limit, not only the temperature required to reach the BEC regime is 
experimentally achievable, but also the theory becomes simple since the Fermi-Fermi mixtures 
can be described effectively by a weakly interacting mixture of molecular bosons and 
excess fermions~\cite{pieri,iskin-mixture,iskin-mixture2,taylor}.
However, the initial proposals of such effective Bose-Fermi mixtures~\cite{pieri,iskin-mixture}
can provide only semi-quantitative results for comparison with experiments in the
BEC regime since the scattering parameters for two molecular bosons and a molecular boson 
and an excess fermion are obtained only in the Born approximation.
In order to overcome this defficiency, we discuss next the boson-fermion 
and boson-boson scattering parameters beyond the Born approximation for arbitrary mass ratio.   
The correct scattering parameters will be used in Sec.~\ref{sec:phase.diagram} 
to construct phase diagrams and density profiles for 
quantitative comparisons with experiments in the BEC regime.

\section{Boson-Fermion and Boson-Boson Scattering Lengths}
\label{sec:a_BF}

Mixtures of two types of fermions in the strong attraction limit can be described 
by effective Bose-Fermi models~\cite{pieri,iskin-mixture,iskin-mixture2,taylor}, 
where fermion pairs behave as molecular bosons and interact weakly with
each other and with excess unpaired fermions.
Scattering lengths between two molecular bosons ($a_{BB}$), and between
a molecular boson and an excess fermion ($a_{BF})$ were calculated
in the Born approximation using many body techniques for 
equal~\cite{carlos, pieri} and unequal masses~\cite{iskin-mixture, iskin-mixture2}.
However, these results do not agree with calculations using 
few body techniques~\cite{petrov,petrov-abf},
because it is necessary to go beyond the Born approximation.

In the case of fermion mixtures with equal masses, while the Born approximation 
in many body theory leads to $a_{BB} = 2a_F$~\cite{carlos} and 
$a_{BF} = 8a_F/3$~\cite{pieri, iskin-mixture}, the results from few body techniques are 
$a_{BB} \approx 0.60a_F$~\cite{petrov} and $a_{BF} \approx 1.18a_F$~\cite{skorniakov}.
However, a diagrammatic approach beyond the Born approximation~\cite{brodsky, levinsen} 
for equal mass fermions recovers the few body results. 
In this section, we generalize these diagrammatic 
approaches and analyze the boson-fermion and boson-boson scattering parameters 
for two types of fermions with unequal masses in order 
to make quantitative predictions for experiments 
involving Fermi-Fermi mixtures in the strong attraction limit. 
Here, we show that the diagrammatic many body approach 
for unequal mass fermions produce results consistent with three and
four body techniques that were recently used to obtain $a_{BF}$ and $a_{BB}$ as a function of 
mass ratio~\cite{petrov, petrov-abf}. 
We analyze the technical aspects of the boson-fermion scattering parameter for unequal
mass fermions in some detail, since they are much easier to present, 
while we do not discuss in great detail the technical aspects of the boson-boson scattering
parameter for unequal masses, as they are extremely cumbersome.
Detailed descriptions of the boson-fermion and boson-boson scattering parameters for equal
mass fermions can be found in the literature~\cite{brodsky, levinsen}.

\begin{figure} [htb]
\psfrag{a}{\LARGE $-\mathbf{k}, w_F$}
\psfrag{b}{\LARGE $\mathbf{k}, w_B + \epsilon_b$}
\psfrag{c}{\LARGE $\mathbf{k}+\mathbf{p}, w_B - w_F + \epsilon_b + p_0$}
\psfrag{d}{\LARGE $-\mathbf{q}, w_F - q_0$}
\psfrag{i}{\LARGE $\mathbf{q}, w_B + \epsilon_b + q_0$}
\psfrag{f}{\LARGE $\mathbf{q}+\mathbf{p}, w_B - w_F + \epsilon_b + p_0 + q_0$}
\psfrag{g}{\LARGE $-\mathbf{p}, w_F - p_0$}
\psfrag{h}{\LARGE $\mathbf{p}, w_B + \epsilon_b + p_0$}
\centerline{\scalebox{0.4}{\includegraphics{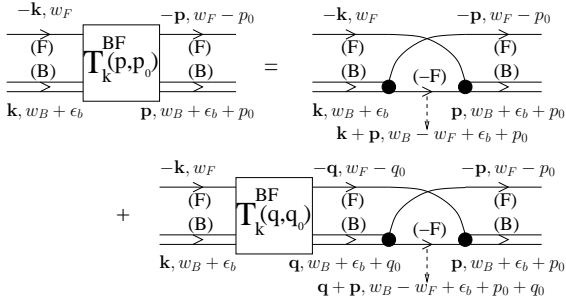}}}
\caption{\label{fig:t-matrix}
Diagrammatic representation of the integral equation for the
boson-fermion scattering T-matrix $T_\mathbf{k}^{BF}(\mathbf{p},p_0)$.
Here, we use the notation $(-\uparrow) \equiv \downarrow$ and vice versa.
}
\end{figure}

We begin our analysis by describing a zero temperature $(T = 0)$ diagrammatic representation for 
the boson-fermion scattering T-matrix $T_\mathbf{k}^{BF}(\mathbf{p},p_0)$ 
as shown in Fig.~\ref{fig:t-matrix}, where $w_F = k^2/(2m_F)$ and $w_B = k^2/(2m_B)$ 
are the kinetic energies for the excess fermions and molecular bosons, respectively, 
and $\epsilon_b = - 1/(m_{\uparrow \downarrow} a_F^2) < 0$
is the binding energy of the molecular bosons.
Here,
\begin{eqnarray}
m_B &=& m_\uparrow + m_\downarrow \\
m_{\uparrow\downarrow} &=& \frac{2m_\uparrow m_\downarrow}{m_\uparrow + m_\downarrow}
\end{eqnarray}
are masses of the molecular bosons and twice the reduced mass of 
the $\uparrow$ and $\downarrow$ fermions, respectively.
In this figure, single lines represent retarded free fermion propagators
\begin{equation}
G_{0,\sigma}(\mathbf{k}, w) = \frac{1}{w - w_\sigma + \mu_\sigma + i0^+},
\end{equation}
where $w_\sigma = k^2/(2m_\sigma)$ is the energy and $\mu_\sigma$ is the
chemical potential of the $\sigma$-type fermions.
Similarly double lines represent the retarded molecular boson propagators 
\begin{equation}
\label{eqn:bose-propagator}
D_0(\mathbf{k}, w) = \frac{4\pi/m_{\uparrow\downarrow}^{3/2}}
{|\epsilon_b|^{1/2} - (w_B - w - \mu_\uparrow - \mu_\downarrow - i0^+)^{1/2}}
\end{equation}
obtained from the expansion of the effective action~\cite{iskin-mixture,iskin-mixture2},
and a corresponding RPA re-summation of the fermion polarization bubbles leading to
$
D_0(\mathbf{k}, w) =  - g/[1 + g \Gamma (\mathbf{k}, w)]
$
where the fermion polarization bubble is 
$
\Gamma (\mathbf{k}, w) = \sum_{{\bf q}, q_0} G_{0,\sigma}(\mathbf{k} + \mathbf{q}, w + q_0 ) 
G_{0,\sigma}(-\mathbf{q}, -q_0 ).
$
Integration over the internal momentum ${\bf q}$ and frequency $q_0$ leads to 
$
\Gamma (\mathbf{k}, w) = 
\Gamma (0,0) + [m_{\uparrow\downarrow}^{3/2} / (4 \pi)]
\left( w_B - w  - \mu_{\uparrow} - \mu_{\downarrow} - i0^{+} \right)^{1/2}
$
which in combination with the definition of the fermion-fermion scattering 
length $a_F = m_{\uparrow\downarrow} T^{FF} (0,0)/ 4\pi$, and
the fermion-fermion T-matrix 
$
T^{FF} (0,0) = - g / [1 + g \Gamma (\mathbf{0}, 0)]
$
lead to the final result described in Eq.~(\ref{eqn:bose-propagator}).

On the right hand side of Fig.~\ref{fig:t-matrix}, the first diagram 
represents a fermion exchange process, and all other (infinitely many) 
possible processes are included in the second diagram.
In all diagrams, we choose $\uparrow$ ($\downarrow$) to label
lighter (heavier) fermions such that lighter (heavier) fermions
are in excess when $F \equiv \uparrow$ $(F \equiv \downarrow)$. 
This choice spans all possible mass ratios.
In the following, we set $\mu_\sigma = 0$ since all of the calculations are 
performed for three fermions scattering in vacuum.
The T-matrix $T_\mathbf{k}^{BF}(\mathbf{p},p_0)$ 
satisfies the following integral equation
\begin{eqnarray}
&& T_\mathbf{k}^{BF}(\mathbf{p},p_0) = 
- G_{0,-F}(\mathbf{k}+\mathbf{p}, w_B - w_F + \epsilon_b + p_0) \nonumber \\
&& - \sum_{\mathbf{q}, q_0} D_0(\mathbf{q}, w_B + \epsilon_b + q_0) 
G_{0,F}(-\mathbf{q}, w_F - q_0) \times \\
&& T_\mathbf{k}^{BF}(\mathbf{q},q_0) G_{0,-F}(\mathbf{p} + \mathbf{q}, w_B - w_F + \epsilon_b + p_0 + q_0) \nonumber,
\label{eqn:t-matrix}
\end{eqnarray}
where we used $(-\uparrow) \equiv \downarrow$ and vice versa.
On the right hand side, we can sum over frequency $q_0$ by closing the 
integration contour in the upper half-plane, where $T_\mathbf{k}^{BF}(\mathbf{q},q_0)$ 
and $D_0(\mathbf{q}, w_B + \epsilon_b + q_0)$ are analytic functions of $q_0$.
Since this integration sets $q_0 = k^2/(2m_F) - q^2/(2m_F)$, we set 
$p_0 = k^2/(2m_F) - p^2/(2m_F)$ in order to have the same frequency dependence
for the T-matrix on both sides~\cite{levinsen}. 
Since we are interested in the zero-range low energy s-wave scattering,
we average out the angular dependences of $\mathbf{k}$ and $\mathbf{p}$.
When $k \to 0$, the generalized integral equation defined in Eq.~(\ref{eqn:t-matrix})
can be expressed in terms of the boson-fermion scattering function $a_{k \to 0} (p)$ as
\begin{eqnarray}
&& \frac{m_{\uparrow\downarrow} a_0^{BF}(p)/m_{BF}} 
{a_F^{-1} + (m_{\uparrow\downarrow} p^2/m_{BF} + a_F^{-2})^{1/2}} 
= \frac{1}{p^2 + a_F^{-2}} - \frac{m_B}{2\pi m_F} \nonumber \\
&& \int_0^\infty \frac{dq}{qp} 
\ln\left( \frac{q^2 + 2m_F qp/m_B + p^2 + a_F^{-2}} {q^2 - 2m_F qp/m_B + p^2 + a_F^{-2}} \right) 
a_0^{BF}(q).
\label{eqn:ieqn}
\end{eqnarray}
Here, we used the definition of the boson-fermion scattering length 
\begin{equation}
a_k^{BF}(p) = \frac{m_{BF}}{m_{\uparrow\downarrow}^{3/2}}
\left[|\epsilon_b|^{1/2} + \left(\frac{p^2-k^2}{m_{BF}} - \epsilon_b\right)^{1/2}\right] 
T_k^{BF}(p),
\label{eqn:a-scattering}
\end{equation}
with its full momentum dependence,
where $m_{BF}$ is twice the reduced mass of an excess fermion 
and a molecular boson given by
\begin{equation}
m_{BF} = \frac{2m_B m_F}{m_B + m_F}.
\end{equation}
The integral equation shown in Eq.~(\ref{eqn:ieqn}) as well
as the scattering length expression shown in Eq.~(\ref{eqn:a-scattering}) 
reduce to the results for the equal masses~\cite{brodsky, levinsen} when 
$m_\uparrow = m_\downarrow = m$.
Notice that only the fermion exchange process is taken into account in 
the Born approximation, and that neglecting the second term on the right hand side 
of Eq.~(\ref{eqn:ieqn}) leads to
$
a_0^{BF}(0) = 2(m_{BF}/m_{\uparrow\downarrow})a_F
$
which is consistent with our previous results~\cite{iskin-mixture, iskin-mixture2}. 
However, we need to include both terms in order to find the exact boson-fermion scattering length.

\begin{figure} [htb]
\centerline{\scalebox{0.5}{\includegraphics{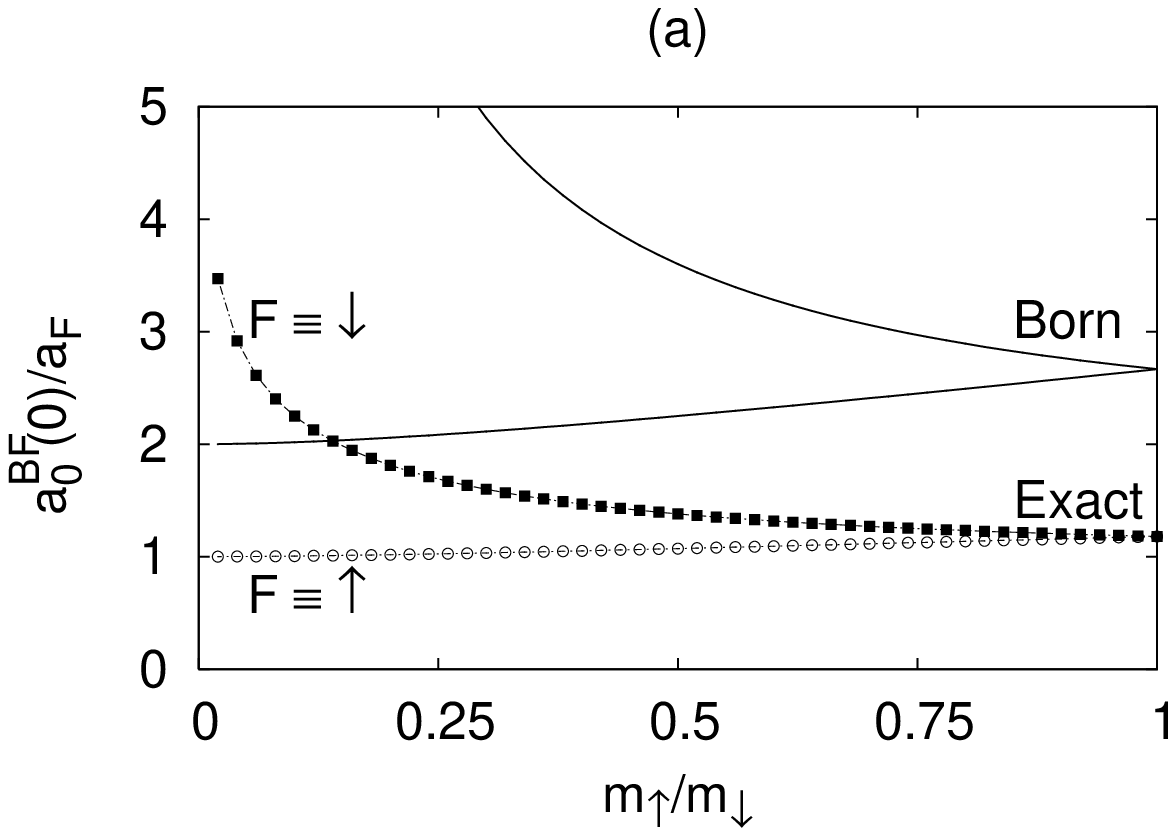}}}
\centerline{\scalebox{0.5}{\includegraphics{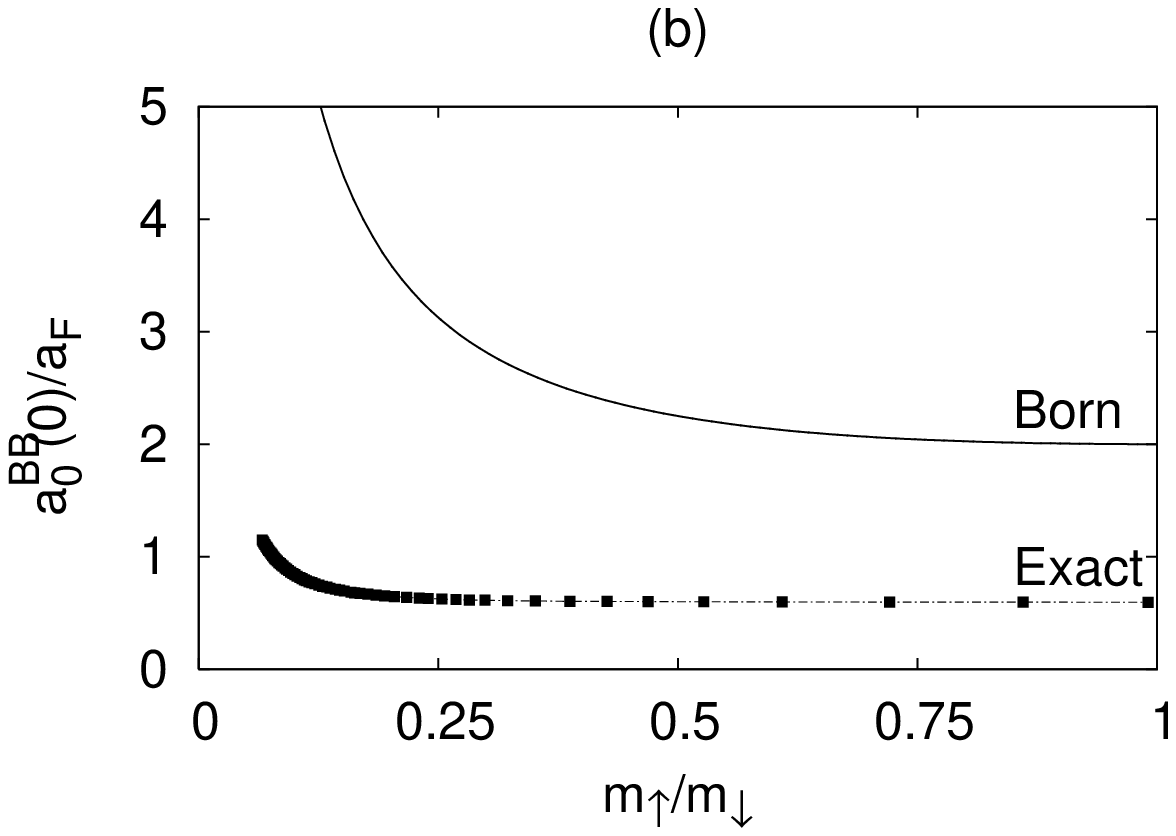}}}
\caption{\label{fig:a_BF-mr}
The boson-fermion scattering length $a_0^{BF}(0)/a_F$ versus mass anisotropy 
$m_\uparrow/m_\downarrow$ is shown in (a) when lighter $\uparrow$-type (hollow circles) or 
heavier (solid circles) $\downarrow$-type fermions are in excess.
The boson-boson scattering length $a^{BB}(0)/a_F$ versus mass anisotropy 
$m_\uparrow/m_\downarrow$ is shown in (b). In addition, the scattering lengths in the 
Born approximation are shown as solid lines.
}
\end{figure}

Next, we solve numerically the integral equation given in Eq.~(\ref{eqn:ieqn}),
and obtain $a_0^{BF}(p)$ as a function of the mass anisotropy $m_\uparrow/m_\downarrow$.
The exact solutions and the Born approximation values of
$a_0^{BF}(0)$ are shown in Fig.~\ref{fig:a_BF-mr}(a).
When $m_\uparrow = m_\downarrow$, we find $a_0^{BF}(0) \approx 1.179a_F$, which 
is consistent with results previously found for equal masses~\cite{skorniakov, petrov, petrov-abf, brodsky, levinsen, taylor}.
Notice that $a_0^{BF}(0)$ decreases (increases) from this value with increasing mass 
anisotropy when the lighter (heavier) fermions are in excess.
In addition, in the limit of $m_\uparrow/m_\downarrow \to 0$, while
$a_0^{BF}(0) \to a_F$ when the lighter fermions are in excess, 
$a_0^{BF}(0)$ grows rapidly when the heavier fermions are in excess.
Notice also that the Born approximation values for $a_0^{BF}(0)$ are not in agreement
with the exact values for any mass anisotropy, but the general qualitative trends 
are captured by the Born approximation as can be seen from Fig.~\ref{fig:a_BF-mr}(a).

In addition, we present results for $a^{BB}(0)$
as a function of mass anisotropy $m_\uparrow/m_\downarrow$ in Fig.~\ref{fig:a_BF-mr}(b).
The exact results of the boson-boson scattering parameters for unequal mass fermions 
can be obtained either by extending the diagrammatic approach
for equal mass fermions~\cite{brodsky, levinsen}
or by using few body techniques~\cite{petrov}. The Born approximation values
$
a^{BB}(0) = (m_B/m_{\uparrow\downarrow}) a_F
$
are found in Ref.~\cite{iskin-mixture, iskin-mixture2}.
Since the technical details to calculate the boson-boson scattering parameters 
are quite cumbersome, and not particularly illuminating, here we just mention  
that the results found in the literature~\cite{petrov} can also be 
obtained diagramatically for the unequal mass case.
As shown in Fig.~\ref{fig:a_BF-mr}(b), $a^{BB}(0)$ grows very slowly as
the mass ratio $m_{\uparrow}/m_{\downarrow}$ decreases, in constrast with
the much more rapid growth of the Born approximation values of $a^{BB}(0)$.
As expected, the Born approximation values of $a^{BB}(0)$ 
are not in agreement with the exact values for any mass anisotropy.

\begin{table} [htb]
\begin{tabular}{|c|c|c|c|c|c|}
\hline
$\uparrow$ & $\downarrow$ & $m_\uparrow/m_\downarrow$ & $a_{BB}/a_F$ & $a_{B\uparrow}/a_F$ & $a_{B\downarrow}/a_F$ \\
\hline
$^{6}$Li & $^{6}$Li & 1.000 & 0.593 & 1.179 & 1.179 \\
$^{6}$Li & $^{40}$K & 0.150 & 0.695 & 1.010 & 1.984 \\
$^{6}$Li & $^{87}$Sr & 0.068 & 1.123 & 1.003 & 2.512 \\
$^{6}$Li & $^{171}$Yb & 0.035 & - & 1.001 & 3.023 \\
$^{40}$K & $^{40}$K & 1.000 & 0.593 & 1.179 & 1.179 \\
$^{40}$K & $^{87}$Sr & 0.460 & 0.597 & 1.064 & 1.411 \\
$^{40}$K & $^{171}$Yb & 0.234 & 0.629 & 1.022 & 1.723 \\
$^{87}$Sr & $^{87}$Sr & 1.000 & 0.593 & 1.179 & 1.179 \\
$^{87}$Sr & $^{171}$Yb & 0.508 & 0.599 & 1.073 & 1.374 \\
$^{171}$Yb & $^{171}$Yb & 1.000 & 0.593 & 1.179 & 1.179 \\
\hline
\end{tabular}
\caption{\label{table:a} Exact boson-boson ($a_{BB}$) and boson-fermion ($a_{BF}$) 
scattering lengths for a list of two-species Fermi-Fermi mixtures.
Here, $a_{B \uparrow}$ ($a_{B \downarrow}$) corresponds to excess-type 
of $\uparrow$ ($\downarrow$) fermions.
}
\end{table}

Several atomic gases of fermions ($^{6}$Li, $^{40}$K, $^{87}$Sr~\cite{innsbruck}, 
and $^{171}$Yb~\cite{fukuhara}) are being currently investigated,
and experimental methods for studying mixtures of two types of fermions 
are being developed in several groups. Thus, anticipating future experiments 
involving mixtures of two types of fermions, 
we show in Table~\ref{table:a} the boson-fermion and boson-boson scattering lengths
for a few mixtures.

Here, we would like to make two comments. First, it is quite remarkable that
the diagramatic many body approach recovers the few body results for boson-fermion and
boson-boson scattering lengths for arbitrary mass ratio $m_{\uparrow}/m_{\downarrow}$.
The diagramatic many body approach is performed in momentum space, while the few body 
approach is performed in real space. The two methods are equivalent because all of the possible
scattering processes are taken into account exactly in the diagramatic many body approach 
at $T = 0$ for three or four fermions. However, the calculation of the 
scattering parameters for three or four fermions in the presence of many others 
(arbitrary number of particles) at zero or finite temperatures requires 
a full many body approach, which is inevitably approximate and more difficult to implement.
Second, our analysis does not include the effects related to 
Efimov (three body bound) states,
which may become important when analyzing the scattering parameters as a function
of mass ratio $m_{\uparrow}/m_{\downarrow}$~\cite{petrov-abf}. In particular, the mixtures of 
$^6$Li and $^{87}$Sr or $^6$Li and $^{171}$Yb have mass ratios of  
$m_{Li}/m_{Sr} \approx 0.068$ and $m_{Li}/m_{Yb} \approx 0.035$, which are below the critical
ratio $m_{\uparrow}/m_{\downarrow} \approx 0.073$ for the emergence of Efimov states. 

Having presented the boson-fermion and boson-boson scattering lengths
for arbitrary mass ratio $m_{\uparrow}/m_{\downarrow}$, we discuss next the
resulting phase diagrams for Fermi-Fermi mixtures in the strong attraction
limit, where the system can be effectively described by a
Bose-Fermi mixture~\cite{pieri,iskin-mixture,iskin-mixture2,taylor}
of molecular bosons and excess unpaired fermions.

\section{Bose-Fermi Mixtures at Zero Temperature}
\label{sec:BFmixture}

In this section we use the effective Bose-Fermi mixture 
description~\cite{pieri,iskin-mixture,taylor}
to analyze the phase diagram of population imbalanced 
Fermi-Fermi mixtures in the strong attraction limit. 
We describe first the general stability conditions 
for Bose-Fermi mixtures at zero temperature, and use 
this connection to discuss the stability and phase 
diagrams of Fermi-Fermi mixtures in the 
strong attraction limit.

\subsection{Bose-Fermi Mixtures}
\label{sec:atomic-BFmixture}

The ground state of Bose-Fermi mixtures can be described by the
free energy~\cite{viverit, iskin-mixture2}
\begin{equation}
{\cal E} = {\cal E}_B + {\cal E}_F + \frac{U_{BB} n_B^2}{2} + U_{BF} n_F n_B 
- \mu_F n_F - \mu_{B} n_B,
\label{eqn:energy}
\end{equation}
which characterizes the center-of-mass degrees of freedom for a mixture
of single-hyperfine-state bosons and fermions. 
Here $\mu_F$ and $n_F$ ($\mu_B$ and $n_B$) are the density and chemical 
potential of fermions (bosons), $\epsilon_F$ is the Fermi energy 
of the fermions, and $U_{BB}$ and $U_{BF}$ are the repulsive 
boson-boson and boson-fermion interaction strengths.
The density of single-hyperfine-state fermions in three dimensions is given by
$
n_F = (1/V) \sum_\mathbf{k}^{|\mathbf{k}| < k_F} 1 = k_F^3/(6\pi^2),
$
where $k_F$ is the Fermi momentum and $V$ is the volume. 
The first term in Eq.~(\ref{eqn:energy}) is the total kinetic energy of 
bosons, which is assumed to be much smaller than all other energies, and is neglected.
This assumption is very good since essentially all bosons are condensed in
the ${\bf k} = {\bf 0}$ state, when the boson-boson and boson-fermion 
interactions are weak.
The second term in Eq.~(\ref{eqn:energy}) is the total kinetic energy of 
fermions, which in three dimensions is given by
$
{\cal E}_F  = (1/V) \sum_\mathbf{k}^{|\mathbf{k}| < k_F} \epsilon_{\mathbf{k},F} 
= 3 \epsilon_F n_F / 5,
$
where $\epsilon_{\mathbf{k},F} = |\mathbf{k}|^2/(2m_F)$.

From the free energy given in Eq.~(\ref{eqn:energy}), we obtain the fermion
and boson chemical potentials using the condition 
$\partial {\cal E} /\partial n_{i} = 0$ with $i = F, B$, leading to
\begin{eqnarray}
\mu_F &=& \epsilon_F + U_{BF} n_B, \\
\mu_B &=& U_{BB} n_B + U_{BF} n_F.
\end{eqnarray}
Then, we use the positive definiteness of the Bose-Fermi compressibility matrix 
$\kappa_{i,j} = \partial \mu_i/\partial n_j$,
\begin{eqnarray}
\frac{\partial \mu_F}{\partial n_F} \frac{\partial \mu_B}{\partial n_B} - 
\frac{\partial \mu_F}{\partial n_B} \frac{\partial \mu_B}{\partial n_F} > 0,
\end{eqnarray}
and find that bosons and fermions phase separate when the condition
\begin{equation}
n_F \ge \frac{4\pi^4}{3m_F^3} \frac{U_{BB}^3}{U_{BF}^6}, \hspace{3mm} \textrm{(3D)}
\label{eqn:stab.3D}
\end{equation}
is satisfied in three-dimensional systems~\cite{viverit, iskin-mixture2}.
Therefore, the stability of uniform superfluidity puts an upper 
limit on the density of fermions in three-dimensions.

Following a similar approach in lower dimensions, where
$n_F = k_F^2/(4\pi)$ and ${\cal E}_F = \epsilon_F n_F / 2$ in two dimensions,
and $n_F = k_F/\pi$ and ${\cal E}_F = \epsilon_F n_F / 3$ in one dimension,
we find that the bosons and the fermions phase separate when the conditions
\begin{eqnarray}
1 &\le& \frac{m_F}{2\pi} \frac{U_{BF}^2}{U_{BB}}, \hspace{3mm} \textrm{(2D)} 
\label{eqn:stab.2D} \\
n_F &\le& \frac{m_F}{\pi^2} \frac{U_{BF}^2}{U_{BB}}, \hspace{3mm} \textrm{(1D)}
\label{eqn:stab.1D}
\end{eqnarray}
are satisfied, respectively, for two- and one-dimensional systems.
Notice that, the stability of uniform superfluidity puts a lower limit
in one dimension, which is in sharp contrast with the three-dimensional result.
Furthermore, the stability condition in two dimensions does not depend
explicitly on the density of fermions (see also Ref.~\cite{tempere}). 
However, the results in lower dimensions have to be used with caution, 
since quantum fluctuations are more pronounced, and may affect these stability conditions. 

For an atomic Bose-Fermi mixture, we can also describe analytically a finer
structure of phases. There are four possible phases~\cite{viverit}:
(I) PS(1) where there is phase separation between pure fermions and pure bosons;
(II) PS(2) where there is phase separation between pure fermions,
and a mixture of fermions and bosons;
(III) PS(3) where there is phase separation between pure bosons,
and a mixture of fermions and bosons; and
(IV) PS(4) where there is  phase separation between two different mixtures 
of fermions and bosons.

For a three-dimensional weakly interacting Bose-Fermi mixture, we follow Ref.~\cite{viverit} 
and find that there are only two stable phases within the phase separation region:
(I) PS(1) where there is phase separation between pure fermions and pure bosons, and 
(II) PS(2) where there is phase separation between pure fermions,
and a mixture of fermions and bosons.
We obtain analytically the condition
\begin{equation}
n_F \ge \frac{1125 \pi^4}{128 m_F^4} \frac{U_{BB}^3}{U_{BF}^6} 
- \frac{5}{4} \frac{U_{BB}}{U_{BF}} n_B, \hspace{3mm} \textrm{(3D)}
\label{eqn:stab2.3D}
\end{equation}
for the transition from the PS(2) to the PS(1) phase~\cite{iskin-mixture2}.

In lower dimensions, we find that the structure of the phase diagram is quite different.
In two dimensions, the phase separated region consists only of 
PS(1) where there is phase separation between pure fermions and pure bosons.
While, for a one-dimensional weakly interacting Bose-Fermi mixture, 
the phase separated region consists also of two regions:
(I) PS(1) where there is phase separation between pure fermions and pure bosons,
and (III) PS(3) where there is phase separation between pure bosons,
and a mixture of fermions and bosons. 
We obtain analytically the condition
\begin{equation}
n_F \le \frac{3 m_F}{2\pi^2} \frac{U_{BF}^2}{U_{BB}} 
- \frac{U_{BB}}{U_{BF}} n_B, \hspace{3mm} \textrm{(1D)}
\label{eqn:stab2.1D}
\end{equation}
for the transition from the PS(3) to the PS(1) phase.
Notice that the structure of the PS(3) phase in one-dimension is 
very different from the structure of the PS(2) phase in three dimensions.
Again, the results in lower dimensions have to be used with caution, 
since the quantum fluctuations are more pronounced, and may affect these stability conditions. 

Next, we concentrate only on the three-dimensional case, and use
the stability conditions found above as well as the interaction (scattering)
parameters obtained in Sec.~\ref{sec:a_BF} to analyze the phase
diagrams of Fermi-Fermi mixtures in the strong attraction limit.

\subsection{Fermi-Fermi Mixtures \\ in the Strong Attraction Limit}
\label{sec:FFmixture}

To make an analogy between Bose-Fermi mixtures and population
imbalanced Fermi-Fermi mixtures in the strong attraction limit, 
we identify $F \equiv \{\uparrow$ or $\downarrow\}$ as the excess fermions.
This identification leads to the density of excess fermions ($n_E$) and molecular
bosons ($n_B$) given by
\begin{eqnarray}
n_E &=& n_F - n_{-F} = |n_\uparrow - n_\downarrow|, \\
n_B &=& \frac{n - n_E}{2} = n_{-F},
\end{eqnarray}
respectively, where $n = n_\uparrow + n_\downarrow$ is the total density of 
$\uparrow$- and $\downarrow$-type fermions.
Here, we use $(-\uparrow) \equiv \downarrow$ and vice versa.
For instance, if $F \equiv \uparrow$ fermions are in excess, the density
of excess fermions and molecular bosons are $n_E = n_\uparrow - n_\downarrow$ and 
$n_B = (n-n_E)/2 = n_\downarrow$, respectively, such that all $\downarrow$-type
fermions are paired with some of the $\uparrow$-type fermions to form molecular bosons,
but there are $\uparrow$-type fermions left unpaired.
It is important to emphasize that the internal degrees of freedom 
(electronic, vibrational, and rotational) of molecular bosons are
not explicitly considered here, in the same spirit of the description of atomic bosons 
presented in Sec.~\ref{sec:atomic-BFmixture}, where the electronic degrees of freedom were
also not explicitly considered. 

For three dimensions, we define the boson-boson and boson-fermion interaction strengths 
\begin{eqnarray}
U_{BB} &=& \frac{4\pi a_{BB}}{m_B} = \frac{4\pi \gamma_B}{m_B} a_F, 
\label{eqn:U_BB} \\
U_{BF} &=& \frac{4\pi a_{BF}}{m_{BF}} = \frac{4\pi \beta_F}{m_{BF}} a_F,
\label{eqn:U_BF}
\end{eqnarray}
where $a_F$, $a_{BB} = \gamma_B a_F$ and $a_{BF} = \beta_F a_F$ are the 
fermion-fermion, boson-boson and boson-fermion scattering lengths.
Here, $\gamma_B = a_{BB}/a_F$ and $\beta_F = a_{BF}/a_F$ are constants, 
which are found in Sec.~\ref{sec:a_BF}
as shown in Fig.~\ref{fig:a_BF-mr} and Table~\ref{table:a}.
In addition, we define the population imbalance parameter
\begin{equation}
P = \frac{N_\uparrow - N_\downarrow}{N_\uparrow + N_\downarrow} 
= \frac{n_\uparrow - n_\downarrow}{n_\uparrow + n_\downarrow},
\end{equation}
such that $|P| = n_E/n$, and $n = K_F^3/(3\pi^2)$, where $N_\sigma$ is the
number of $\sigma$-type fermions and $K_F$ is the Fermi 
momentum corresponding to the total density of fermions defined
by $K_F^3 = (k_{F,\downarrow}^3 + k_{F,\uparrow}^3)/2$.

Using these definitions, the phase separation condition Eq.~(\ref{eqn:stab.3D}) becomes
\begin{equation}
|P| \ge \frac{\pi^3 \gamma_B^3 m_{BF}^6}{16 \beta_F^6 m_B^3 m_F^3} \lambda^3,
\label{eqn:psc}
\end{equation}
where $\lambda = 1/(K_F a_F)$ is the scattering parameter.
Similarly, the condition given in Eq.~(\ref{eqn:stab2.3D}) becomes
\begin{equation}
|P| \left(1 - \frac{5 \gamma_B m_{BF}}{8 \beta_F m_B} \right) 
\ge \frac{3375 \pi^3 \gamma_B^3 m_{BF}^6}{8192 \beta_F^6 m_B^3 m_F^3}\lambda^3 
- \frac{5\gamma_B m_{BF}}{8 \beta_F m_B},
\label{eqn:psc2}
\end{equation}
for the transition from the PS(2) to the PS(1) phase.

We would like to emphasize that these analytic expressions given in Eqs.~(\ref{eqn:psc}) 
and~(\ref{eqn:psc2}) are valid only in the strong attraction limit 
when $1/(K_F a_F) \gg 1$, but  
they may still give semi-quantitative results for $1/(K_F a_F) \gtrsim 1$.
Close to the unitarity, the Bose-Fermi description of Fermi-Fermi mixtures 
in terms of molecular bosons and excess fermions is not reliable, since the binding energy 
of molecular bosons is small and the interactions between molecular bosons and excess fermions 
or between two molecular bosons may be sufficient to cause dissociation of the
molecules into directly scattering fermions. 
However, there may be an intermediate regime between unitarity and the 
strict BEC limit where we can describe Fermi-Fermi mixtures in terms
of a mixture of molecular bosons and excess fermions such that
the molecular bosons can dissociate due to boson-boson or boson-fermion interactions,
but be in chemical equilibrium with excess fermions. 
When dissociation of molecular bosons is included, the system is no longer
a binary mixture of molecular bosons and excess fermions, but a 
ternary mixture of molecular bosons, dissociated bosons ($\uparrow \downarrow \rightleftharpoons \uparrow + \downarrow$), 
and excess fermions, or effectively a ternary mixture of molecular bosons, 
and $\uparrow$- and $\downarrow$-type fermions. 
In the case of ternary mixtures, there can be a large number of phase separated regimes. 
If we confine our discussion to the equillibrium of a maximum of two phases 
of this ternary mixture then several other situations can be encountered. 
For example, when $\downarrow$-type fermions are in excess, a possible sequence of phases
for fixed population imbalance $P$ and increasing scattering parameter $1/(K_F a_F)$ is:
(1) Normal phase (N) of partially polarized fermions $\to$ 
(2) mixture of molecular bosons
and $\uparrow$-type fermions phase separated from partially polarized normal 
fermions $\to$ 
(3) molecular bosons phase separated from excess $\downarrow$-type
fermions $\to$ 
(4) mixture of molecular bosons and $\downarrow$-type excess fermions
phase separated from $\downarrow$-type excess fermions $\to$ 
(5) coexistence of molecular bosons and $\downarrow$-type excess fermions.

Therefore, as long as Fermi-Fermi mixtures can be regarded as a binary mixture
of non-dissociated molecular bosons and excess fermions,  
the expressions given in Eqs.~(\ref{eqn:psc}) and~(\ref{eqn:psc2}) 
may be used as a guide for the boundaries between phase 
separated (non-uniform) and the mixed (uniform) phases for any mixture of fermions. 
In particular, Eqs.~(\ref{eqn:psc}) and~(\ref{eqn:psc2}) may serve as estimators for the phase boundaries
of equal or unequal mass Fermi-Fermi mixtures with population imbalance, 
as discussed next.

\subsection{Phase Diagrams of Fermi-Fermi Mixtures in the Strong Attraction Limit}
\label{sec:phase.diagram}

Among many possibilities of Fermi-Fermi mixtures (see Table~\ref{table:a}), we focus our
analysis on population imbalanced mixtures of 
$^{6}$Li or $^{40}$K atoms where $m_\uparrow = m_\downarrow$, and
$^{6}$Li and $^{40}$K atoms where $m_\uparrow = 0.15 m_\downarrow$.

\begin{figure} [htb]
\centerline{\scalebox{0.5}{\includegraphics{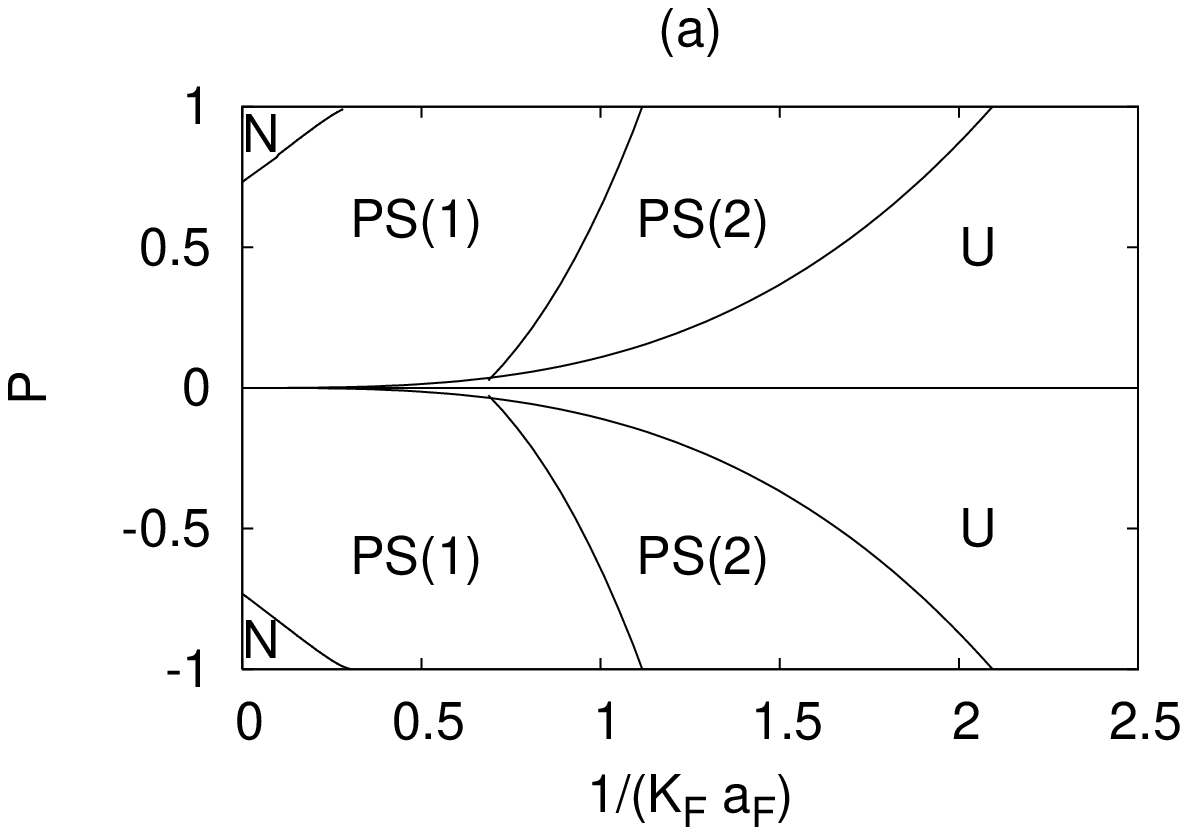} }}
\centerline{\scalebox{0.5}{\includegraphics{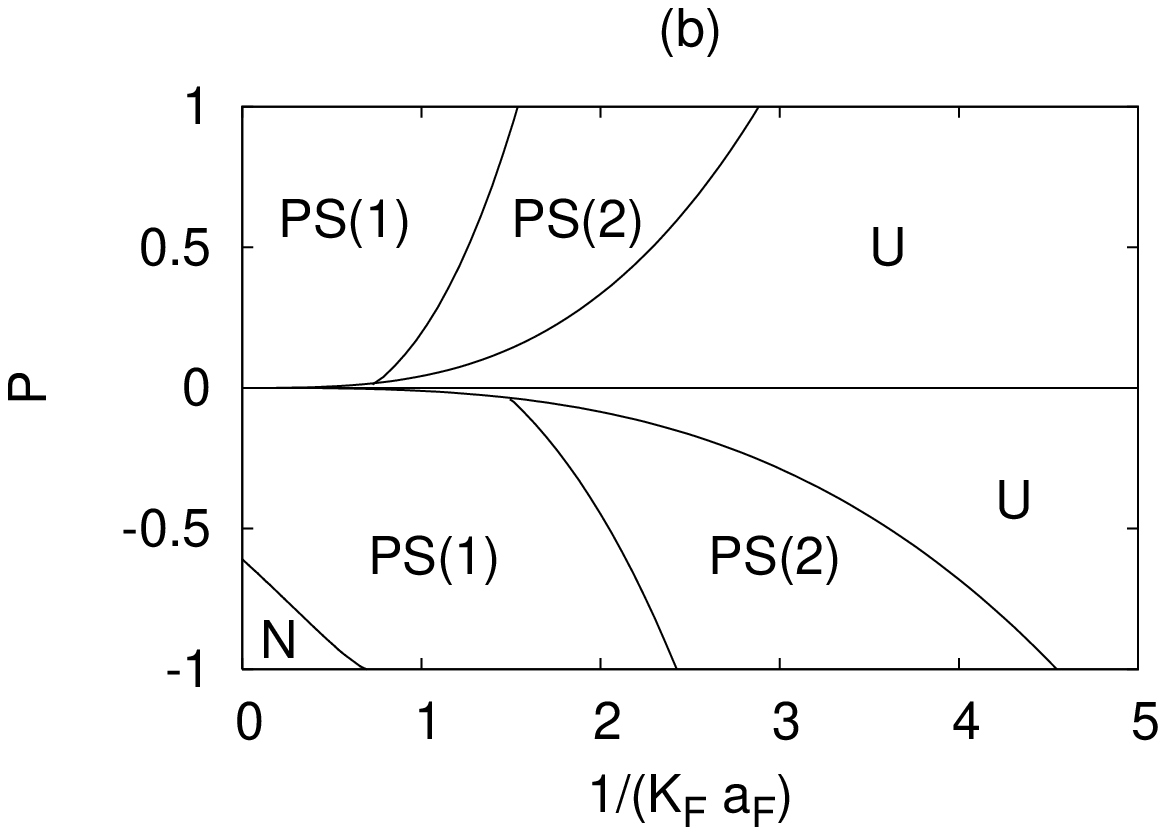} }}
\caption{\label{fig:BEC}
Phase diagram of population imbalance $P = (n_\uparrow - n_\downarrow)/(n_\uparrow + n_\downarrow)$ 
versus scattering parameter $1/(K_F a_F)$ for 
(a) equal masses when $m_\uparrow = m_\downarrow$, and
(b) unequal masses when $m_\uparrow = 0.15 m_\downarrow$
in the BEC side.
We show normal (N), uniform superfluid (U),
and phase separated non-uniform superfluid phases PS(1) and PS(2).
}
\end{figure}

In Fig.~\ref{fig:BEC}, we show phase diagrams of population imbalance $P$
and scattering parameter $1/(K_F a_F)$ for equal mass mixtures 
when $m_\uparrow = m_\downarrow$, and for unequal mass mixtures 
when $m_\uparrow = 0.15 m_\downarrow$.
In these diagrams, we choose $\uparrow$ ($\downarrow$) to label
lighter (heavier) fermions such that lighter (heavier) fermions
are in excess when $P > 0$ $(P < 0)$. 
Although these diagrams are strictly valid in the strong attraction limit
when $1/(K_F a_F) \gg  1$, they are qualitatively correct 
when $1/(K_F a_F) \gtrsim  1$ or as long as the molecular bosons are
not dissociated. In the later case, the system may be approximately 
described as a ternary mixture of molecular bosons,
$\uparrow$- and $\downarrow$-type fermions and many other phases are 
possible, as discussed in Sec.~\ref{sec:FFmixture}.

In these figures, we show the following phases: 
(i) the normal (N) phase corresponding to balanced $(P = 0$) or 
imbalanced $(P \ne 0)$ mixture of unpaired $\uparrow$- or $\downarrow$-type fermions;
(ii) uniform superfluid (U) phase where paired
(molecular bosons) and unpaired fermions coexist; 
and (iii) phase separated (PS) non-uniform superfluid phases.
The PS(1) region labels phase separation between pure excess fermions and 
superfluid molecular bosons, while the PS(2) region labels phase separation between 
pure excess fermions, and a mixture of excess fermions and superfluid molecular bosons.
The phase boundary between U and PS(2) phases is determined from Eq.~(\ref{eqn:psc}), 
and the phase boundary between PS(2) and PS(1) phases is determined 
from Eq.~(\ref{eqn:psc2}).
For a fixed mass anisotropy, when $|P|$ is large, 
we find phase transitions from PS(1) $\to$ PS(2) $\to$ U as 
the interaction strength $1/(K_F a_F)$ increases.
However, when $|P|$ is very small, we find a phase transition directly from the 
PS(1) to the U phase as $1/(K_F a_F)$ increases.

We would like to remark in passing that the phase diagrams for 
mixtures of $^6$Li and $^{87}$Sr or $^6$Li and $^{171}$Yb with mass ratios of  
$m_{Li}/m_{Sr} \approx 0.068$ and $m_{Li}/m_{Yb} \approx 0.035$, which are below the critical
ratio $m_{\uparrow}/m_{\downarrow} \approx 0.073$ for the emergence 
of Efimov (three body bound) states are much richer, since phase separation and
coexistence phases involving Efimov states (trimers), molecular bosons
and excess fermions are also present.

\begin{figure} [htb]
\centerline{\scalebox{0.5}{\includegraphics{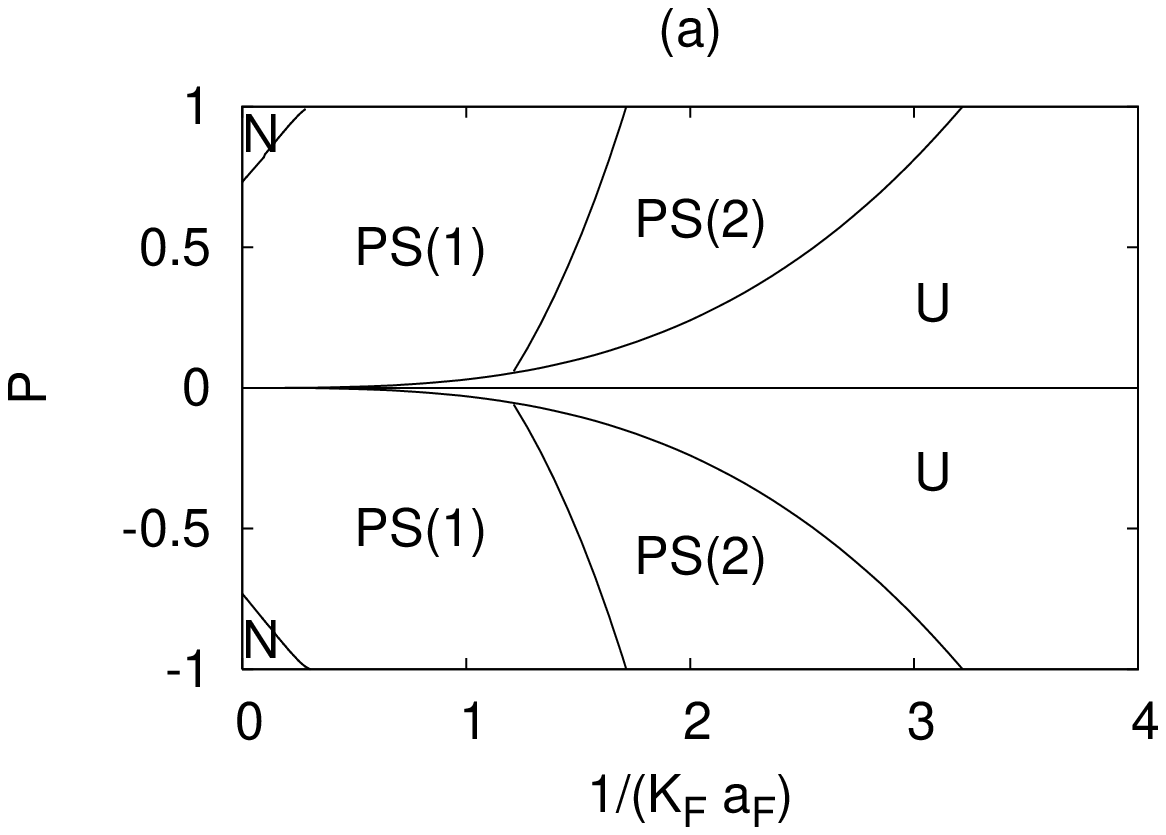} }}
\centerline{\scalebox{0.5}{\includegraphics{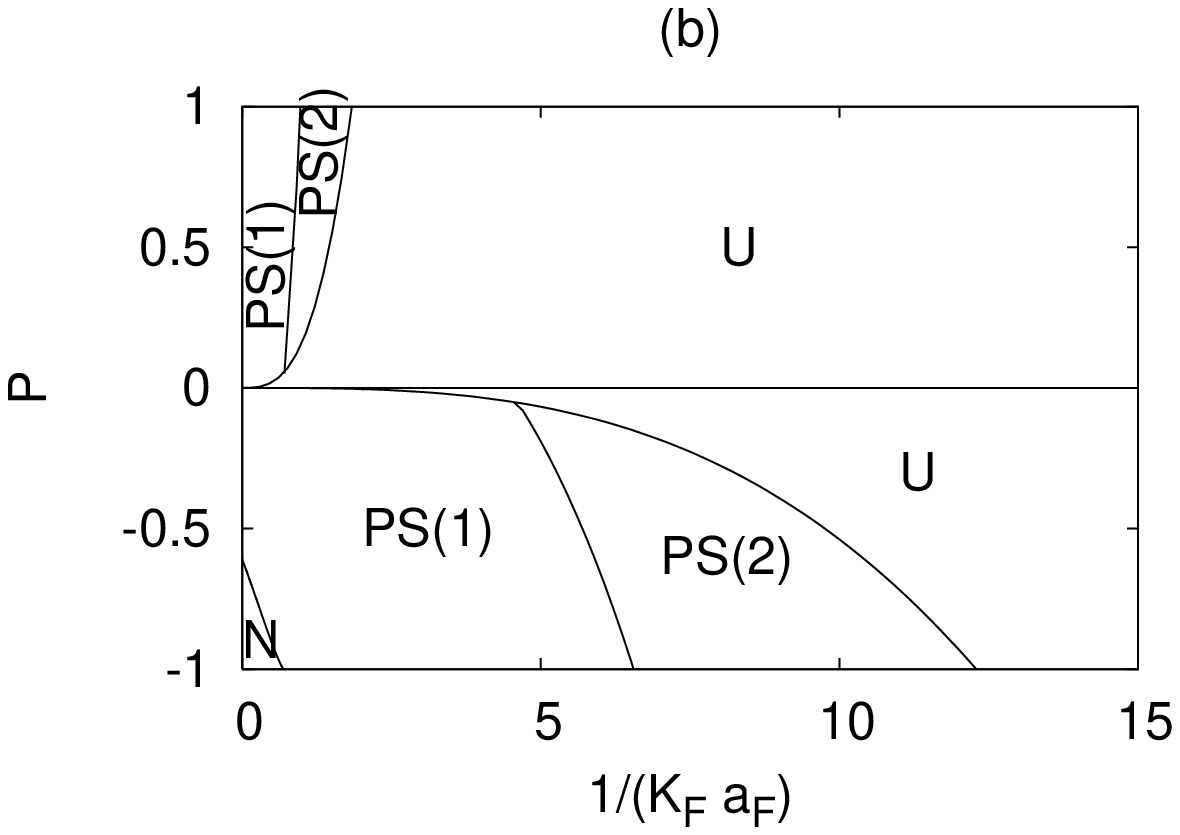} }}
\caption{\label{fig:BEC-born}
Phase diagram of population imbalance $P = (n_\uparrow - n_\downarrow)/(n_\uparrow + n_\downarrow)$ 
versus scattering parameter $1/(K_F a_F)$ in the Born approximation for 
(a) equal masses when $m_\uparrow = m_\downarrow$, and
(b) unequal masses when $m_\uparrow = 0.15 m_\downarrow$
in the BEC side.
We show normal (N), uniform superfluid (U),
and phase separated non-uniform superfluid phases PS(1) and PS(2).
}
\end{figure}

It is also important to emphasize that since we use the exact boson-boson and 
boson-fermion scattering lengths, our phase diagrams in the strong 
attraction limit already include fluctuation corrections beyond the Born approximation. 
For comparison, the corresponding phase diagrams within the Born approximation 
are described in Fig.~\ref{fig:BEC-born}, where the phase boundaries in the
population imbalance $P$ versus scattering parameter $1/(K_F a_F)$ plane are shown
for equal $(m_\uparrow = m_\downarrow)$ and unequal $(m_\uparrow = 0.15 m_\downarrow)$
mass mixtures. A direct comparison of Figs.~\ref{fig:BEC} and~\ref{fig:BEC-born}
shows that the results beyond the Born approximation are quantitatively 
different from the saddle-point results~\cite{pao, sheehy} in the equal mass case.
These quantitative differences become significantly large for unequal 
mass mixtures when heavier fermions are 
in excess~\cite{iskin-mixture,iskin-mixture2,pao-mixture,parish} 
due to the large sensitivity of the exact scattering parameters on 
the mass ratio $m_{\uparrow}/ m_{\downarrow}$ as shown
in Fig.~\ref{fig:a_BF-mr} and Table~\ref{table:a}.
However, the same phases are present in both cases, indicating that the 
Born approximation captures the basic qualitative features, but fails
to produce the phase boundaries quantitatively.

Lastly, we would like to point out the presence of several triple points
in the phase diagrams shown in Figs.~\ref{fig:BEC} and~\ref{fig:BEC-born}. 
Along the $\vert P \vert = 1$ lines, we find several triple points as $1/(K_F a_F)$ increases
where the fully polarized normal phases ($P = \pm 1$) merge with (i) the partially polarized normal (N)
and the PS(1) phase; or with (ii) the PS(1) and PS(2) phases; or with (iii) 
the PS(2) and U phases. Furthermore, there is also an additional triple point that
occurs for small $\vert P \vert$ (iv) where the phases PS(1), PS(2) and U meet.
The precise locations of these triple points can be obtained for any mass ratio 
and scattering parameter from Eqs.~(\ref{eqn:psc})
and~(\ref{eqn:psc2}) using the equal sign $(=)$ condition.
The triple point for case (ii) can be obtained by setting $\vert P \vert = 1$ in 
Eq.~(\ref{eqn:psc2}), and the triple point for case (iii) 
can be obtained by setting $\vert P \vert = 1$ in Eq.~(\ref{eqn:psc}). 
Finally, the triple point for case (iv) can be obtained
by using the equal sign $(=)$ condition of Eqs.~(\ref{eqn:psc}) and~(\ref{eqn:psc2}) and by
solving the two equations simultaneously.

Having analyzed the phase diagrams for non-trapped continuous systems, we discuss
next the effects of the trapping potential which 
are necessary to understand experiments involving ultracold 
Fermi-Fermi mixtures.

\section{Trapped Bose-Fermi Mixtures at Zero and Finite Temperatures}
\label{sec:trap}

In this section, we use again the simpler description of the effective Bose-Fermi mixture
to describe trapped Fermi-Fermi mixtures in the strong attraction limit.
For this purpose, we present first the theory of trapped Bose-Fermi mixtures 
at zero and finite temperatures, and then discuss the density profiles 
of trapped Fermi-Fermi mixtures in the strong attraction limit using the
relation between the two systems described in Sec.~\ref{sec:BFmixture}.

\subsection{Bose-Fermi mixtures}
\label{sec:trap.atomic}

The Hamiltonian density for a Bose-Fermi mixture in an external potential can be 
written as
\begin{eqnarray}
\label{eqn:bose-fermi-hamiltonian}
H ( {\bf r} ) & = & K_F( {\bf r} ) + V_F( {\bf r} ){\hat n_F} ( {\bf r} ) + K_B( {\bf r} ) 
+ V_B( {\bf r} ) {\hat n_B} ( {\bf r} ) \nonumber \\
& + & U_{BF} {\hat n_B} ( {\bf r} ) {\hat n_F} ( {\bf r} )  
+ (U_{BB}/2) {\hat n_B} ( {\bf r} ) {\hat n_B}( {\bf r} ), 
\end{eqnarray}
where $ K_i( {\bf r} ) = 
\psi_{i}^{\dagger} ( {\bf r} ) [-\nabla^2/(2m_i) - \mu_i] \psi_{i} ( {\bf r} )$ 
represents the kinetic and chemical potential terms for fermions $(i \equiv F)$ 
or bosons $(i \equiv B)$ in a single hyperfine state. 
Here, ${\hat n_i}({\bf r}) = \psi_{i}^{\dagger}({\bf r}) \psi_i ({\bf r})$
represent the local density operators, and $U_{BF}$ and $U_{BB}$ represent the 
boson-fermion and boson-boson interaction. The single-hyperfine-state fermions
are non-interacting, but obey the Pauli exclusion principle. 

For simplicity, we approximate the trapping potential by an isotropic harmonic function
where the potential energy is 
\begin{eqnarray}
V_{i} ( {\bf r} ) = \frac{1}{2} \alpha_{i} r^2.
\end{eqnarray}
Here, $\alpha_{i} = m_{i} w_{i}^2$ is proportional to the trapping 
frequency of bosons $(i \equiv B)$ or fermions $(i \equiv F)$, which is typically different 
for each kind of atom. Since the potential and the interactions are isotropic,
the effective potentials and densities depend only on $r = \vert {\bf r} \vert$.

In the presence of such trapping potentials, the bosons and fermions 
feel the effective potentials 
\begin{eqnarray}
V_{B,{\rm eff}} ( r ) &=& V_B ( r ) + 2U_{BB} n_B(r) + U_{BF} n_F(r),
\label{eqn:V_B} \\
V_{F,{\rm eff}} ( r ) &=& V_F (r) + U_{BF} n_B(r),
\label{eqn:V_F}
\end{eqnarray}
respectively, where $n_F (r)$ is the local density of fermions, and 
\begin{equation}
n_B(r) = n_C(r) + n_{NC}(r).
\end{equation}
is the total local density of bosons.
Here, $n_C (r)$ and $n_{NC} (r)$ are the density of condensed and non-condensed 
bosons, respectively.

The number of condensed bosons is determined from the Gross-Pitaevskii  
equation leading to
\begin{equation}
n_C (r) = \frac{\mu_B - V_B (r) - 2U_{BB} n_{NC}(r) - U_{BF} n_F(r)} {U_{BB}}
\label{eqn:n_Cr}
\end{equation}
within the Thomas-Fermi approximation (TFA), where the kinetic energy
of the bosons is neglected.
This relation is valid when the condition 
$ \mu_B - V_B (r) - 2U_{BB} n_{NC} (r) - U_{BF} n_F(r) \ge 0$
is satisfied, otherwise, $ n_C(r) = 0 $. 

In our analysis, we treat both the non-condensed bosons and the fermions 
as ideal gases~\cite{amoruso} in effective potentials 
given in Eqs.~(\ref{eqn:V_B}) and~(\ref{eqn:V_F}).
In addition, we assume that non-condensed bosons are in thermal equilibrium
with condensed bosons at the same chemical potential $\mu_B$.
Within these approximations, the density of non-condensed bosons and 
fermions are given by
\begin{eqnarray}
n_{NC}(r) &=& \frac{1}{V}\sum_{\mathbf{k}} b[\epsilon_{\mathbf{k},B} - \mu_B + V_{B,{\rm eff}}(r)], 
\label{eqn:n_NCr} \\
n_F(r) &=& \frac{1}{V}\sum_{\mathbf{k}} f[\epsilon_{\mathbf{k},F} - \mu_F + V_{F, {\rm eff}}(r)],
\label{eqn:n_Fr}
\end{eqnarray}
where $b(x) = 1/[\exp(x/T) - 1]$ is the Bose,
and $f(x) = 1/[\exp(x/T) + 1]$ is the Fermi distribution.
Here, $\epsilon_{\mathbf{k},i} = |\mathbf{k}|^2/(2m_{i})$ is the kinetic energy
of bosons $(i \equiv B)$ or fermions $(i \equiv F)$.

Notice that, at zero temperature, all bosons condense 
such that $n_{NC}(r) = 0$, and $n_B(r) = n_C(r)$, leading to 
\begin{eqnarray}
n_B(r) &=& \frac{\mu_B - V_B(r) - U_{BF} n_F(r)} {U_{BB}}, \\
n_F(r) &=& \frac{\{2m_F [\mu_F - V_F(r) - U_{BF} n_B(r) ]\}^{3/2} }{6\pi^2}, 
\end{eqnarray}
for the densities of bosons and fermions, respectively~\cite{iskin-mixture2}. 
The first expression is valid when the condition 
$\mu_B - V_B(r) - U_{BF} n_F(r) \ge 0$ is satisfied, 
otherwise, $n_B(r) = 0$.
The second expression is valid when the condition 
$\mu_F - V_F(r) - U_{BF} n_B(r) \ge 0$ is satisfied, 
otherwise, $n_F(r) = 0$.

The chemical potentials of bosons and fermions are determined from 
fixing the total number of bosons and the number of fermions, 
independently, as follows
\begin{eqnarray}
N_B &=& \int d^3r n_B(r), 
\label{eqn:N_B}\\
N_F &=& \int d^3r n_F(r),
\label{eqn:N_F}
\end{eqnarray}
where the integration is over all space. Therefore, in order to find the density 
profiles for condensed and non-condensed bosons, as wells as for fermions, 
we need to solve Eqs.~(\ref{eqn:N_B}) and~(\ref{eqn:N_F}) 
for $\mu_B$ and $\mu_F$ self-consistently.
Next, we discuss the density profiles of Fermi-Fermi mixtures 
in the strong attraction limit using the effective Bose-Fermi description presented.

\subsection{Fermi-Fermi Mixtures \\ in the Strong Attraction Limit}
\label{sec:trap.FF}

To make the connection between Bose-Fermi mixtures and population
imbalanced Fermi-Fermi mixtures in the strong attraction limit,
we identify $F \equiv \{\uparrow$ or $\downarrow\}$ as the excess fermions.
This identification leads to the density of excess fermions ($n_E$) and molecular
bosons ($n_B$) given by
\begin{eqnarray}
n_E(r) &=& n_F(r) - n_{-F}(r) = |n_\uparrow(r) - n_\downarrow(r)|, \\
n_B(r) &=& \frac{n(r) - n_E(r)}{2} = n_{-F}(r),
\end{eqnarray}
respectively, where $n(r) = n_\uparrow(r) + n_\downarrow(r)$ is the total density of 
$\uparrow$- and $\downarrow$-type fermions.
Here, we use $(-\uparrow) \equiv \downarrow$ and vice versa.
For instance, if $F \equiv \uparrow$ fermions are in excess, the density
of excess fermions and molecular bosons are $n_E (r) = n_\uparrow (r) - n_\downarrow (r)$ 
and $n_B(r) = [n (r) - n_E (r)]/2 = n_\downarrow(r)$, respectively, such that 
all $\downarrow$-type fermions are paired with some of the $\uparrow$-type fermions 
to form molecular bosons, but there are $\uparrow$-type fermions left unpaired.
In addition, we identify $\alpha_B = \alpha_\uparrow + \alpha_\downarrow$, where
$\alpha_\sigma = m_\sigma w_\sigma^2$ is different for different atoms.

\begin{figure} [htb]
\centerline{\scalebox{0.49}{\includegraphics{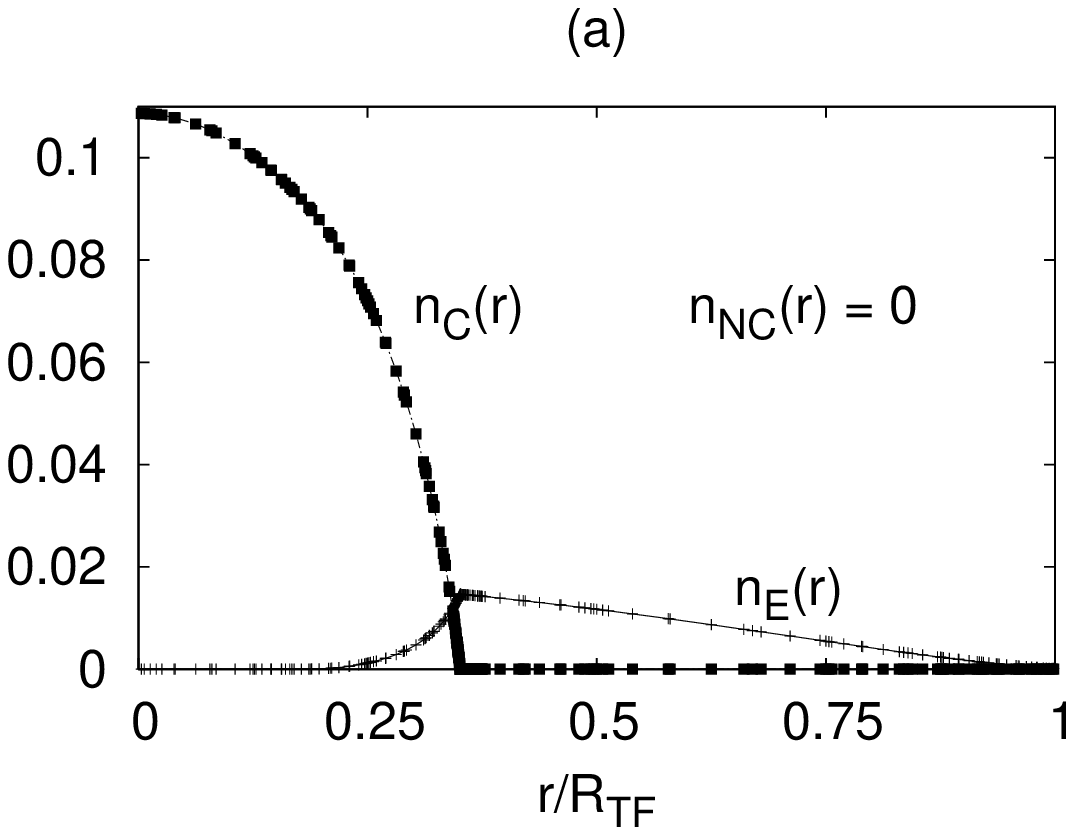} }}
\centerline{\scalebox{0.49}{\includegraphics{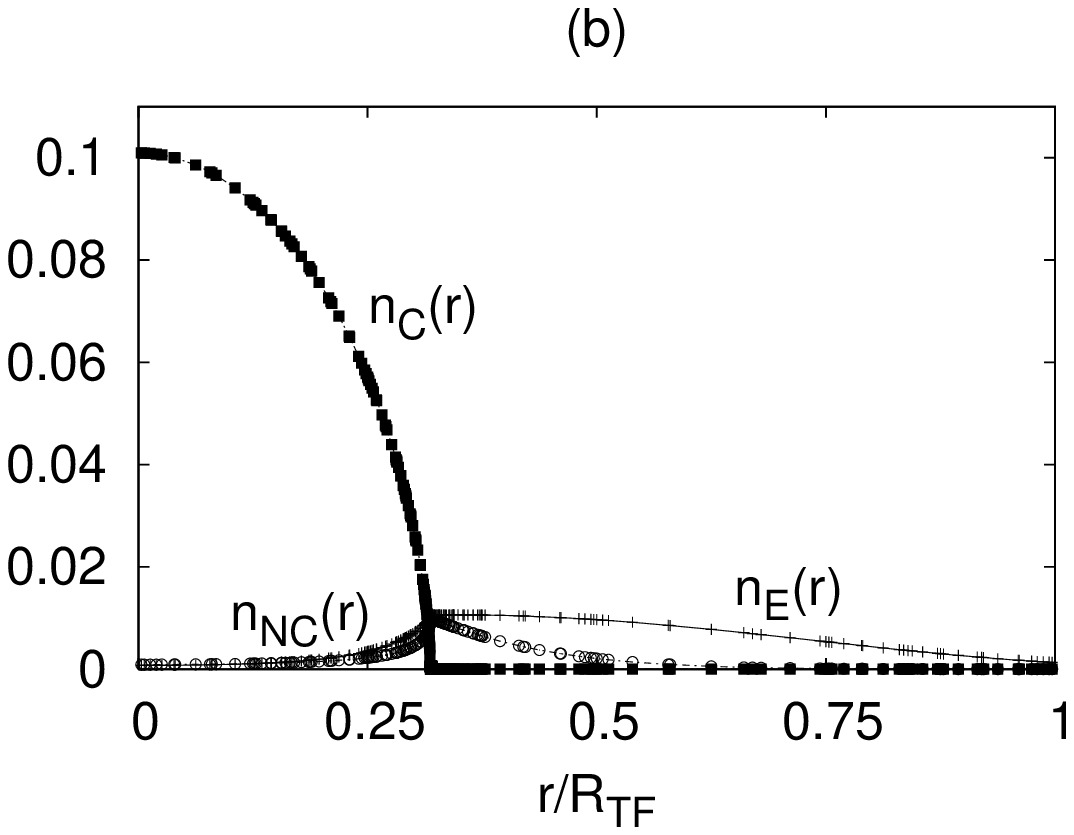} }}
\centerline{\scalebox{0.49}{\includegraphics{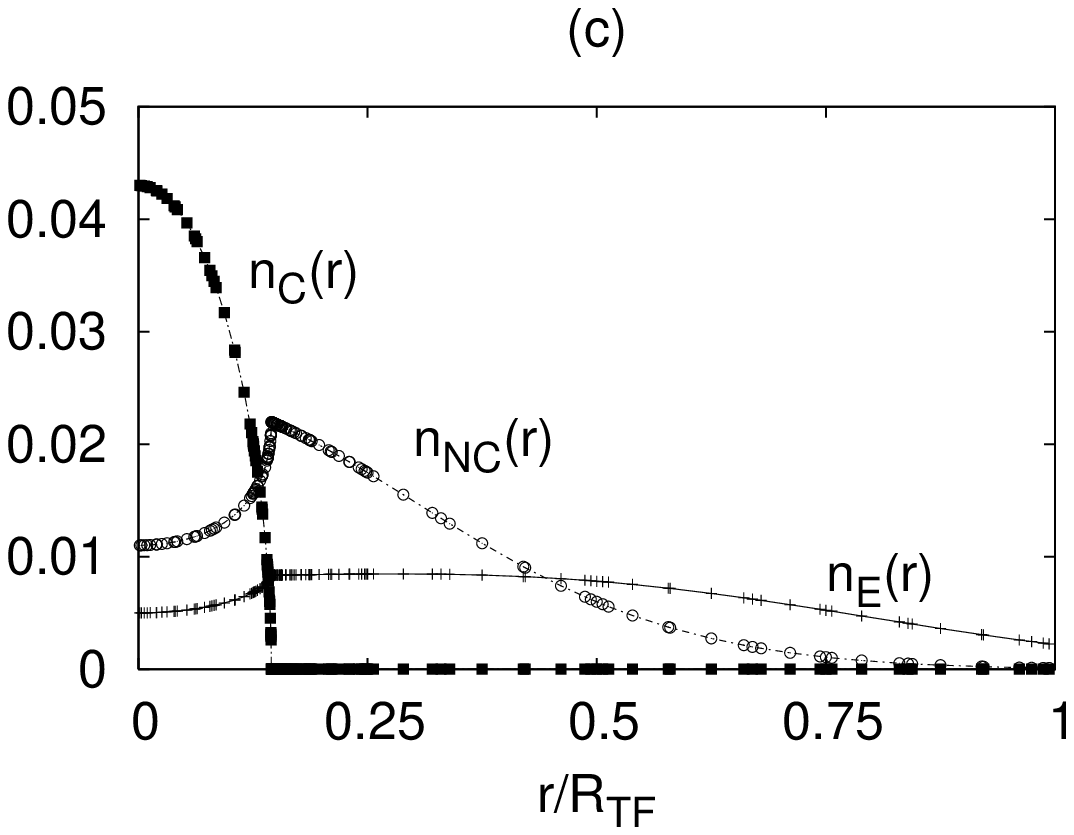} }}
\centerline{\scalebox{0.49}{\includegraphics{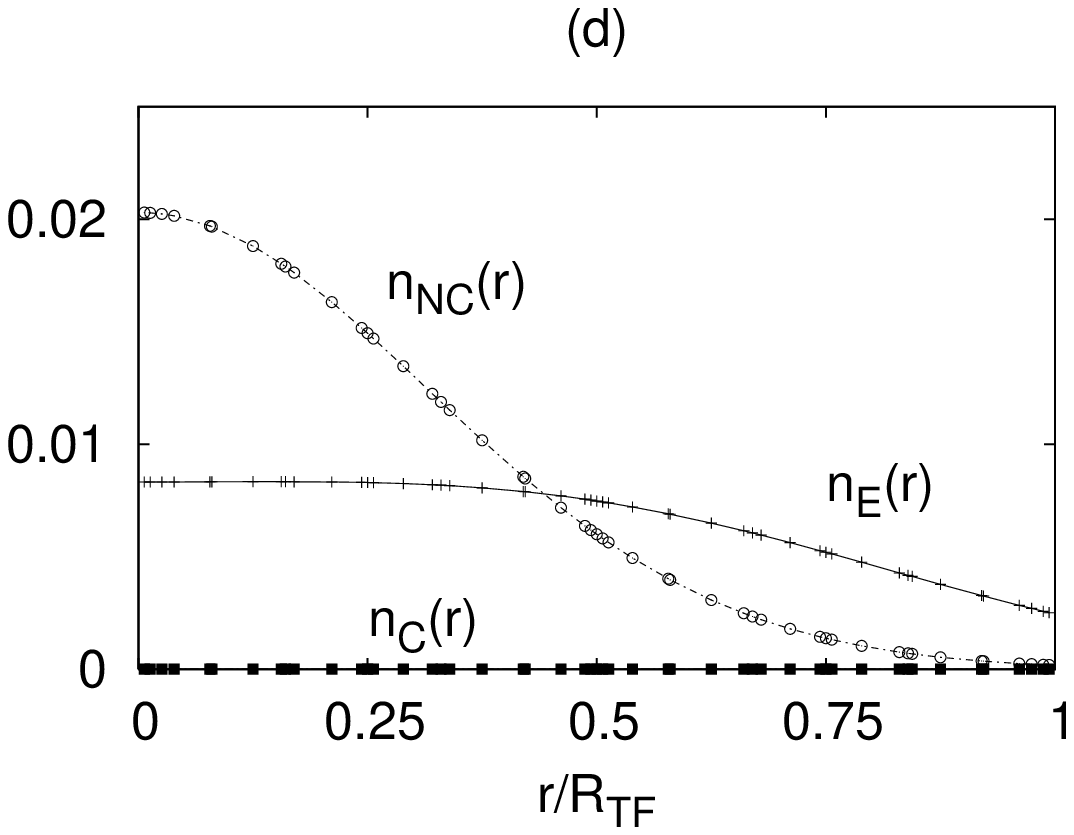} }}
\caption{\label{fig:ncond}
Density (in units of $K_F^3$) of condensed ($n_C$) and non-condensed ($n_{NC}$) 
molecular bosons, and excess ($n_E$) fermions versus trap radius $r/R_{TF}$
is shown for 
(a) $T = 0$,
(b) $T = 0.2 E_F$,
(c) $T = 0.35 E_F$, and
(d) $T = 0.41 E_F$.
Here, $m_\uparrow = m_\downarrow$, the population imbalance parameter is $P = 0.5$ 
and the scattering length parameter is $1/(K_F a_F) = 2$.
}
\end{figure}

As an example, we look at the equal mass case $m_F = m_{\uparrow} = m_{\downarrow} = m$
and $m_B = m_{\uparrow} + m_{\downarrow} = 2m$, and solve the self-consistent relations
Eqs.~(\ref{eqn:N_B}) and~(\ref{eqn:N_F}) for a two-hyperfine-state mixture of $^6$Li only
or $^{40}$K only. In our numerical analysis, we choose 
$\alpha_F = \alpha_\uparrow = \alpha_\downarrow = \alpha$
and $\alpha_B = 2\alpha$, scattering parameter $1/(K_F a_F) = 2$ 
and population imbalance parameter $P = 0.5$ such that $n_\uparrow = 3n_\downarrow$. 
We define a characteristic energy $E_F = K_F^2/(2m)$ in terms of the wavevector $K_F$ 
and the fermion mass $m$ and assume it to be the known Fermi energy of a gas of non
interacting fermions in the trapping potential $V_F(r)$. 
We scale the radial distance with the effective Thomas-Fermi (TF) radius $R_{TF}$ 
defined by $E_F  = \alpha R_{TF}^2/2$. With this identification, the total number of trapped fermions 
$N = (E_F/w_F)^3/3$, with $w_F =  \sqrt{\alpha/m}$, 
can be rewritten as $N = K_F^3 R_{TF}^3/24$ in terms of $K_F$ and $R_{TF}$.  

Since we are interested also in the temperature dependence of the density
profiles, we recall that the critical temperature for Bose-Einstein condensation 
of a non-interacting harmonically trapped Bose gas is $T_{BEC} = w_B [ N_B /\zeta (3) ]^{1/3}$, 
where $w_B = \sqrt{\alpha_B/ m_B}$. In our Fermi-Fermi mixture the number of bosons is
$N_B = N (1 - \vert P \vert )/2$ expressed in terms of the total
number of fermions $N = N_\uparrow + N_\downarrow$ and population imbalance 
$P = (N_\uparrow - N_\downarrow)/N$.
Using the expression for $N_B$, the expression of $N$ in terms of $E_F$ and $w_F$,
and that $w_B = w_F$ for equal masses, we find 
\begin{equation}
T_{BEC} = \left[ \frac{1-|P|}{6 \zeta(3)} \right]^\frac{1}{3} E_F 
\approx 0.518 (1-|P|)^{1/3} E_F,
\label{eqn:tbec}
\end{equation}
for the critical BEC temperature, which is valid when $1/(K_F a_F) \to \infty$. 
Here, $\zeta(x)$ is the Zeta function and $\zeta(3) \approx 1.202$.
Therefore, for $P = 0.5$, we obtain $T_{BEC} \approx 0.41 E_F$.
Notice that $T_{BEC}$ for population imbalance $P$ given in Eq.~(\ref{eqn:tbec}) is a generalization 
of the results for equal populations~\cite{griffin, perali}.

In Fig.~\ref{fig:ncond}, we show the density (in units of $K_F^3$) of condensed 
($n_C$) and non-condensed ($n_{NC}$) molecular bosons, and excess ($n_E$) 
fermions as a function trap radius $r/R_{TF}$ for four temperatures:
(a) $T = 0$,
(b) $T = 0.2 E_F$,
(c) $T = 0.35 E_F$, and
(d) $T = 0.41 E_F$.
At zero temperature $(T = 0)$, as shown in Fig.~\ref{fig:ncond}(a), we find that 
all of the molecular bosons are condensed, and that they are concentrated 
close to the center of the trap.
In contrast, the majority of excess fermions are pushed away from the center 
towards the edges of the trap due to the repulsive boson-fermion interaction and
the high concentration of condensed molecular 
bosons. Therefore, there is a clear indication of phase separation between 
molecular bosons and excess fermions.
When the temperature is increased to $T = 0.2E_F$
shown in Fig.~\ref{fig:ncond}(b), some of the molecular bosons are not 
condensed. These non-condensed molecular bosons are also pushed away
from the center towards the edges of the trap just like the excess fermions.
Further increase in temperature increases (decreases) the number of 
non-condensed (condensed) molecular bosons as can be seen in Fig.~\ref{fig:ncond}(c).
For temperatures close to $T_{BEC}$ and above, all of the molecular bosons 
become non-condensed as shown in Fig.~\ref{fig:ncond}(d) 
having a Gaussian-like density distribution.
Similarly, the excess fermions also have Gaussian-like density distribution 
for temperatures at $T_{BEC}$ and above due to the absence of the condensate. 

Therefore, at zero temperature, we find that the trapping potential tends to 
favor phase separation into a PS(1)-rich phase where regions of almost pure fermions 
and almost pure bosons are separated. However, at finite temperatures, the system
develops a PS(2)-rich phase where regions of almost pure fermions and of almost 
fully mixed bosons and fermions are separated. The region of coexistence of
bosons and fermions can be further broken down into a domain of coexisting
excess fermions with condensed and non-condensed bosons, and into a domain of 
coexisting excess fermions and non-condensed bosons as can be seen in 
Fig.~\ref{fig:ncond}(d). Again, if the molecular bosons are allowed to dissociate,
the system can be described by a ternary 
mixture as discussed in Sec.~\ref{sec:FFmixture} and
the phase diagram can be even richer, especially closer to the unitarity.

\begin{figure} [htb]
\centerline{\scalebox{0.49}{\includegraphics{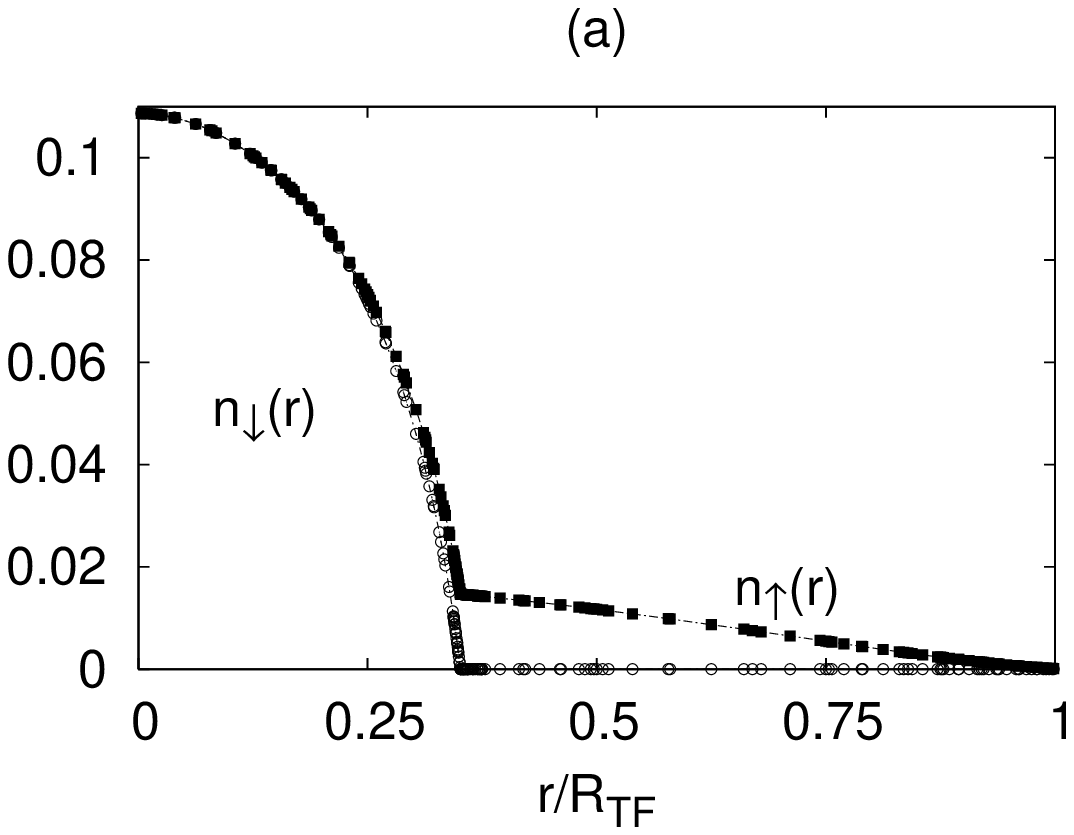} }}
\centerline{\scalebox{0.49}{\includegraphics{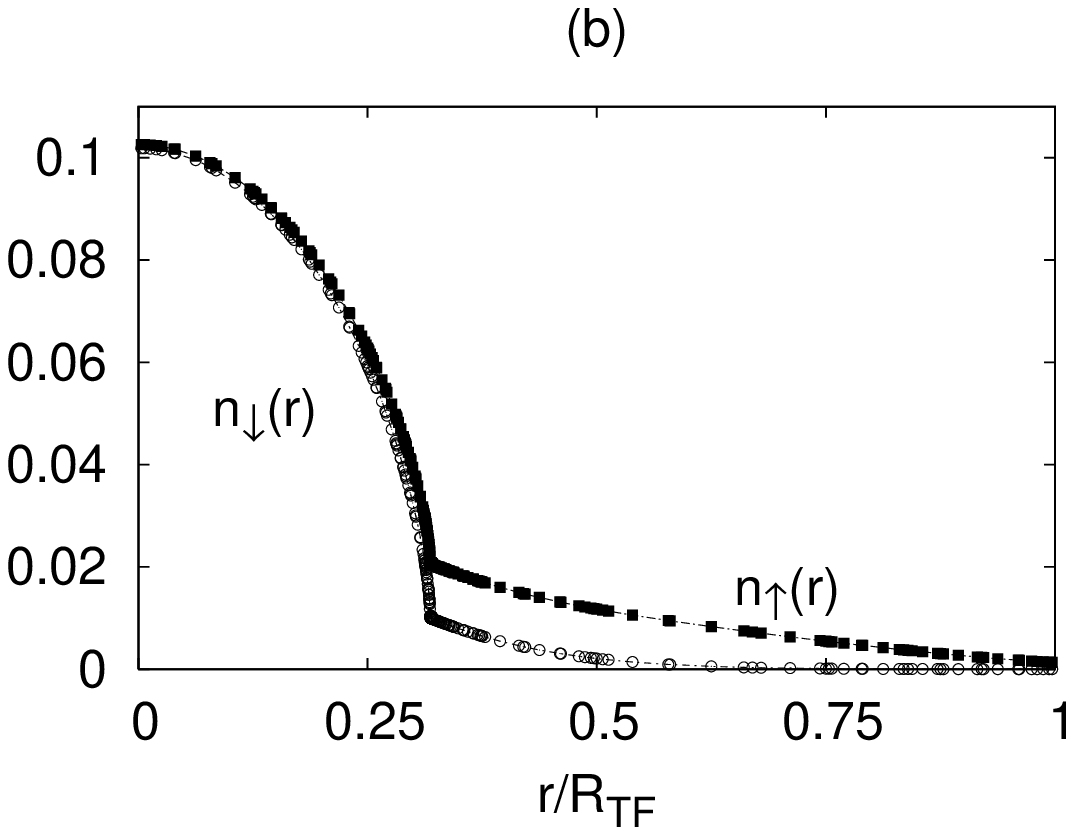} }}
\centerline{\scalebox{0.49}{\includegraphics{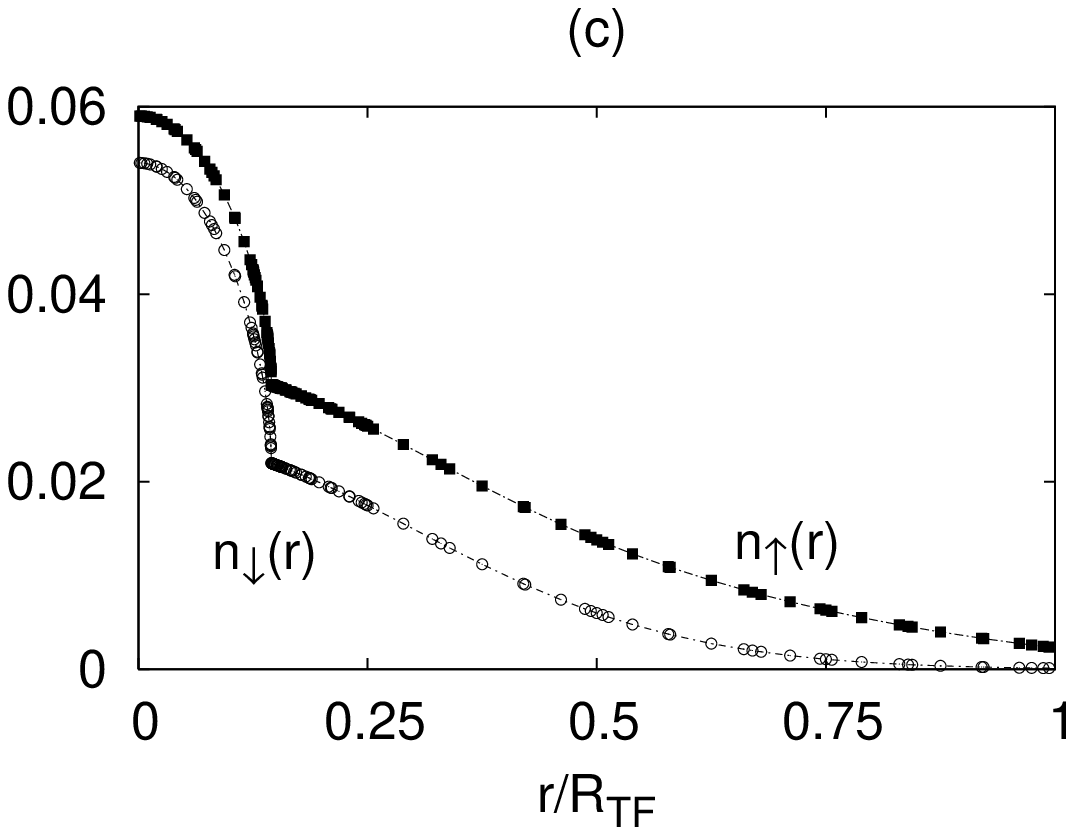} }}
\centerline{\scalebox{0.49}{\includegraphics{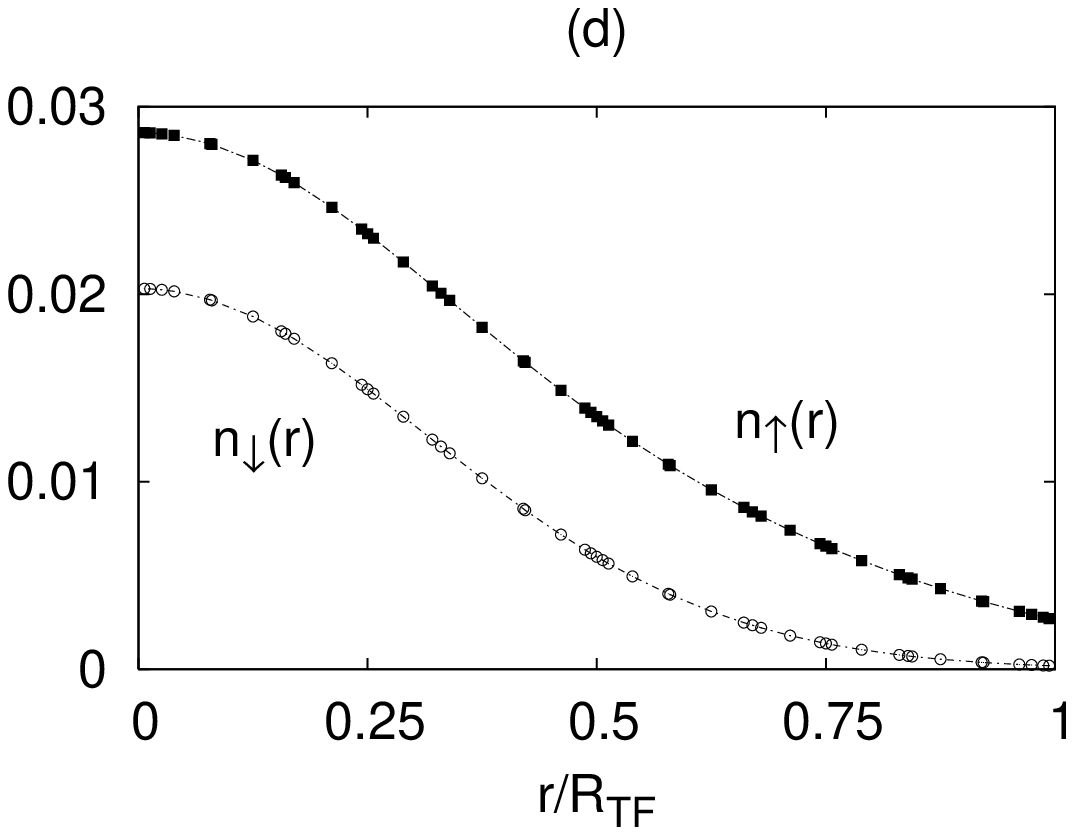} }}
\caption{\label{fig:nupdown}
Density (in units of $K_F^3$) of $\uparrow$- ($n_\uparrow$) and $\downarrow$-type
($n_\downarrow$) fermions versus trap radius $r/R_{TF}$
is shown for 
(a) $T = 0$,
(b) $T = 0.2 E_F$,
(c) $T = 0.35 E_F$, and
(d) $T = 0.41 E_F$.
Here, $m_\uparrow = m_\downarrow$, population imbalance parameter is $P = 0.5$ 
and scattering length parameter is $1/(K_F a_F) = 2$.
}
\end{figure}

In Fig.~\ref{fig:nupdown}, we show the density (in units of $K_F^3$)
of $\uparrow$- ($n_\uparrow$) and $\downarrow$-type ($n_\downarrow$) fermions 
as a function of radius $r/R_{TF}$ for four temperatures:
(a) $T = 0$,
(b) $T = 0.2 E_F$,
(c) $T = 0.35 E_F$, and
(d) $T = 0.41 E_F$.
At zero temperature, as shown in  Fig.~\ref{fig:nupdown}(a), we find that the density 
of $\uparrow$- and $\downarrow$-type fermions are similar close to the center of the 
trapping potential, while some of the excess-type fermions are close to the edges.
When the temperature increases to $T = 0.2E_F$
shown in Fig.~\ref{fig:nupdown}(b) or to $T = 0.35E_F$
shown in Fig.~\ref{fig:nupdown}(c), the density of $\uparrow$- and $\downarrow$-type
fermions become different at the center of the trap. In addition, 
both $\uparrow$- and $\downarrow$-type fermions exist towards the
edges. At temperatures close to $T_{BEC}$ and above, the density profiles 
of $\uparrow$- and $\downarrow$-type fermions have the standard shapes of weakly interacting
trapped Fermi gases.

In this section, we have shown that the effective Bose-Fermi description of Fermi-Fermi mixtures
is applicable in the strong attraction limit, and thus can provide good quantitative
comparisons to experiments in the same regime, since the
exact boson-fermion and boson-boson scattering parameters
were used to obtain the phase diagrams and density profiles.
Thus, we think that it is particularly important to perform experiments
for different population imbalances in the strong attraction limit, 
where the theory is simple. The situation is somewhat more complicated 
near unitarity where quantitative comparisons between theory and experiment 
are more difficult. Furthermore, there are also some differences between the experimental studies
of MIT~\cite{mit, mit-2} and Rice~\cite{rice,rice-2} groups performed near unitarity,
since the shapes of their traps and the number of trapped atoms are quite different.

Having concluded the analysis of the effects of trapping potentials on 
Fermi-Fermi mixtures in the strong attraction limit, next we give a summary of our conclusions.

\section{Conclusions}
\label{sec:conclusions}

In summary, we used the effective Bose-Fermi mixture description to obtain
the phase diagrams of Fermi-Fermi mixtures with equal or 
unequal masses and equal or unequal populations in the strong attraction limit.
For this purpose, we analyzed first the exact boson-fermion 
and boson-boson scattering lengths as a function of mass anisotropy, 
and then we constructed the phase diagrams of Fermi-Fermi mixtures
in the BEC regime.

We showed that three-dimensional  non-trapped fermion mixtures with population imbalance
exhibit phase separation in addition to the normal polarized mixture of fermions 
and uniform mixture of superfluid and excess fermions. 
In the BEC regime, we found two different non-uniform phase separated states: 
PS(1), where there is phase separation between pure unpaired (excess) 
and pure paired fermions (molecular bosons); and
PS(2), where there is phase separation between pure excess fermions 
and a mixture of excess fermions and molecular bosons.
For equal mass mixtures, our results for the phase boundaries
are quantitatively different from previous saddle-point results,
and these quantitative differences become more pronounced 
for unequal mass mixtures when heavier fermions are in excess 
indicating the importance of taking into account scattering processes 
beyond the Born approximation. 

We also discussed the effects of trapping potentials on the density profiles of 
condensed and non-condensed molecular bosons, and excess fermions at zero and finite 
temperatures. At zero temperature, we found that almost all of the condensed
bosons are at the center of the trap, while the excess fermions are pushed to the edges
due to the repulsive boson-fermion interactions. At finite temperatures, we found that
non-condensed pairs and excess fermions are created at the center of the trap 
at the expense of an overall reduction of condensed bosons. 
Finally, at temperatures above the BEC temperature, the number of condensed bosons vanish,
and the system becomes a mixture of weakly interacting non-condensed bosons
and excess fermions. Finally, we discussed that our findings can provide 
good quantitative comparisons to experiments performed in the same 
regime of validity of the theory (BEC regime), since the boson-fermion 
and boson-boson scattering parameters that enter our calculations are exact 
in the dilute limit.

Lastly, we think that it is important to perform experiments with Fermi-Fermi mixtures
in the strong attraction limit (BEC regime) where the theoretical description is
simple. In this limit, additional superfluid and normal phases and richer density 
profiles proposed here can be observed, and directly compared with the theory.

We thank NSF (DMR-0304380) for support.

\end{document}